\documentclass[a4paper,nobibnotes,nofootinbib,notitlepage,superscriptaddress]{revtex4-1}

\usepackage{amssymb,amsmath}
\usepackage{epsfig,subfigure}

\def\b#1{\mathbf{#1}}
\def\D#1{$\displaystyle{#1}$}
\def\f{\frac}
\def\h#1{\widehat{#1}}

\def\lp#1{\left(#1\right)}
\def\o#1{\overline{#1}}
\def\p{\partial}
\def\r{\rightarrow}
\def\tr{\mathsf{Tr\,}}
\def\v#1{\overrightarrow{#1}}
\def\vev#1{\left\langle#1\right\rangle}
\def\FC{\mathcal{F}}

\newcommand{\be}{\begin{equation}}
\newcommand{\ee}{\end{equation}}
\newcommand{\bea}{\begin{eqnarray}}
\newcommand{\eea}{\end{eqnarray}}

\begin{document}

\title{Transverse-momentum-dependent gluon distributions from JIMWLK evolution}

\author{C. Marquet}\email{cyrille.marquet@polytechnique.edu}
\affiliation{Centre de Physique Th\'eorique, \'Ecole Polytechnique, CNRS, Universit\'e Paris-Saclay, F-91128 Palaiseau, France}

\author{E. Petreska}\email{elena.petreska@usc.es}
\affiliation{Centre de Physique Th\'eorique, \'Ecole Polytechnique, CNRS, Universit\'e Paris-Saclay, F-91128 Palaiseau, France}
\affiliation{Departamento de F\'isica de Part\'iculas and IGFAE, Universidade de Santiago de Compostela, 15782 Santiago de Compostela, Spain}

\author{C. Roiesnel}\email{claude.roiesnel@polytechnique.edu}
\affiliation{Centre de Physique Th\'eorique, \'Ecole Polytechnique, CNRS, Universit\'e Paris-Saclay, F-91128 Palaiseau, France}

\begin{abstract}

Transverse-momentum-dependent (TMD) gluon distributions have different operator definitions, depending on the process under consideration. 
We study that aspect of TMD factorization in the small-$x$ limit, for the various unpolarized TMD gluon distributions encountered in the literature.
To do this, we consider di-jet production in hadronic collisions, since this process allows to be exhaustive with respect to the possible operator
definitions, and is suitable to be investigated at small $x$. Indeed, for forward and nearly back-to-back jets, one can apply both the TMD factorization and
Color Glass Condensate (CGC) approaches to compute the di-jet cross-section, and compare the results. Doing so, we show that both descriptions
coincide, and we show how to express the various TMD gluon distributions in terms of CGC correlators of Wilson lines, while keeping $N_c$ finite.
We then proceed to evaluate them by solving the JIMWLK equation numerically. We obtain that at large transverse momentum, the process dependence
essentially disappears, while at small transverse momentum, non-linear saturation effects impact the various TMD gluon distributions in very different ways.
We notice the presence of a geometric scaling regime for all the TMD gluon distributions studied: the "dipole" one, the Weizs\"acker-Williams one, and the
six others involved in forward di-jet production.

\end{abstract}

\maketitle

\section{Introduction}

In hadronic collisions that feature a large transfer of momentum, the standard perturbative QCD framework of collinear factorization is appropriate to calculate scattering cross sections, which are measurable in particular at the Large Hadron Collider. However, some hadronic processes involve, in addition, smaller momentum scales, and for those one needs to resort to a more involved QCD framework, using the concept of transverse-momentum-dependent (TMD) parton distributions, or in short, TMDs. This relates to a large number of observables such as the production of heavy bosons at small transverse momentum \cite{Collins:1984kg,Qiu:2000ga}, transverse spin asymmetries measured in high-energy collisions with polarized beams \cite{Ralston:1979ys,Boer:2003cm}, or in general hadronic scattering in the high-energy limit \cite{Lipatov:1985uk,Catani:1990eg}.

One of the main theoretical obstacles has been the fact that, even in cases for which TMD factorization could be established, the precise operator definition of the parton distributions is dependent on the process under consideration \cite{Mulders:2000sh,Belitsky:2002sm}, implying a loss of universality. In this paper, our goal is to study that aspect of TMDs, in the limit of small longitudinal momentum fraction $x$, where the parton transverse momentum $k_t$ generically plays a central role. Restricting ourselves to unpolarized gluon TMDs, we investigate what happens when the large gluon density reaches the saturation regime, and how the different gluon TMDs are affected by non-linear effects when $k_t$ becomes of the order of the saturation scale $Q_s(x)$, or below. To do so, we shall use the Color Glass Condensate (CGC) framework, an effective theory of QCD which encompasses its small-$x$ dynamics, both in the linear and non-linear regimes \cite{Gelis:2010nm}.

In order to perform our study of the various unpolarized gluon TMDs for protons and nuclei in the small-$x$ regime, we choose to consider the process of forward di-jet production in proton-proton (p+p) and proton-nucleus (p+A) collisions, respectively.
On the one hand, di-jets, when produced nearly back-to-back, provide the two necessary transverse momentum scales, and the strong ordering needed between them, for TMDs to be relevant: the hard scale is the typical single-jet transverse momentum $P_t$ while the softer scale is the total transverse momentum of the jet pair $k_t$, and TMD factorization applies when $k_t\ll P_t$ \cite{Vogelsang:2007jk}. On the other hand, the production at forward rapidities probes small values of $x$: for kinematical reasons, only high-momentum partons from the "projectile" hadron contribute, while on the "target" side, it is mainly small-$x$ gluons that are involved \cite{Marquet:2007vb}. The forward di-jet process is therefore an ideal playground to apply both the TMD and CGC frameworks and to compare them.

Note that the asymmetry of the problem, $x_1\sim 1$ and $x_2\ll 1$, implies that gluons from the target have a much bigger average transverse momentum (of the order of $Q_s$) compared to that of the partons from the projectile (which is of the order of $\Lambda_{QCD}$). Therefore we shall always neglect the transverse momentum of the high-$x_1$ partons from the projectile compared to that of the low-$x_2$ gluons from the target. As a result, the parton content of the projectile hadron will be described by regular parton distributions and TMDs will be involved only on the target side, with the transverse momentum of those small-$x_2$ gluons being equal to the transverse momentum of jet pair $k_t$. This simplification is actually needed in order to apply TMD factorization for the di-jet process, since for this final state, there is no such factorization with TMDs for both incoming hadrons \cite{Collins:2007nk,Rogers:2010dm}.

In order to compare the TMD and CGC approaches in their overlapping domain of validity, we could have considered a simpler process where this issue does not arise, such as for instance semi-inclusive deep-inelastic scattering in electron-proton or electron-nucleus collisions \cite{Ji:2004wu,Marquet:2009ca}. However, with simpler processes, one encounters only small sub-sets of all the possible operator definitions for the gluon TMDs. The advantage of the di-jet process in p+p or p+A collisions is that it involves all the possible gluon TMDs encountered so far in the literature \cite{Bomhof:2006dp}, and therefore it allows us to be comprehensive and study the specifics of the process dependence of gluon TMDs at small-$x$ in an exhaustive way. All our findings, such as the geometric scaling of all the gluon TMDs, will naturally carry over to other processes for which only one or a few of them play a role, like di-jet or heavy-quark production in deep-inelastic scattering and Drell-Yan or photon-jet in p+p and p+A collisions, for instance \cite{Dominguez:2010xd}.

The TMD description of the forward di-jet process, valid in the $k_t\ll P_t$ limit but with no (other than kinematical) constraints on the value of $x_2$, calls for the use of eight different operator definitions for the gluon TMDs \cite{Kotko:2015ura}. They all involve a correlator of two field strength operators, but they differ from each other in their gauge link content. Each of the gluon TMDs is also associated to a different hard factor, made of a sub-set of the possible $2\to2$ diagrams. We show that in the small-$x$ limit, all the gluon TMDs can be simplified and expressed as Fourier transforms of Wilson-line correlators, made either of two, four, six or eight Wilson lines, but with only two different transverse positions whose difference is conjugate to the transverse momentum $k_t$.

The CGC description of the forward di-jet process, valid in the small-$x$ limit but with no (other than kinematical) constraints on the values of the transverse momenta of the jets, involves correlators of up to eight Wilson lines, all of which sit at different transverse positions \cite{Marquet:2007vb,Dominguez:2011wm}. We show that in the $k_t\ll P_t$ limit, the CGC formula coincides with the small-$x$ limit of the TMD formula. In particular, we show how the various gluon TMDs emerge from the framework, how their different operator definitions correspond to different Wilson lines structure of the CGC correlators. We obtain full agreement in the overlapping domain of validity, hereby extending the results of \cite{Dominguez:2011wm} to the case of finite $N_c$.

It is important to note that in the $k_t\ll P_t$ limit, saturation effects do not disappear. Indeed, even though the hard scale $P_t$ is much bigger than the saturation scale, the transverse momentum of jet pair $k_t$ may be of the order of $Q_s$, and formally all powers of $Q_s^2/k_t^2$ may still be included in the definition of the gluon TMDs. They are all contained if the Wilson-line correlators are properly evaluated in the CGC. In particular, the non-linear QCD evolution of all the gluon TMDs can be obtained from the Jalilian-Marian-Iancu-McLerran-Weigert-Leonidov-Kovner (JIMWLK)
\cite{JalilianMarian:1997jx,JalilianMarian:1997dw,Iancu:2000hn,Ferreiro:2001qy,Weigert:2000gi} equation, in the leading $\ln(1/x)$ approximation. Using a numerical simulation of the JIMWLK equation on a discretized lattice, we are able to extract their $k_t$ dependence and evolution towards small values of $x_2$, as well as some important properties: all the gluon TMDs feature geometric scaling (i.e. they are functions of $k_t/Q_s(x_2)$ only, as opposed to $k_t$ and $x_2$ separately) in the saturation region $k_t\leq Q_s(x_2)$, and they either vanish or coincide for $k_t\gg Q_s(x_2)$. Finally, having understood how their process dependence manifests itself in the CGC allows us to restore universality: potential information extracted from a particular process, for one gluon TMD, can be consistently fed into the others.

The paper is organized as follows. In section II, we recall the TMD factorization formula for forward di-jets as well as the operator definitions of the eight gluon TMDs involved, and we explain the simplifications obtained in the small-$x$ limit. In section III, we recall the CGC formula for forward di-jets, take the $k_t\ll P_t$ limit, and show that the result coincides with the one obtained in the TMD framework, in this overlapping domain of validity. In section IV, we describe the numerical method used in order to solve the JIMWLK equation: a lattice implementation of the Langevin formulation of the equation. In section V, we present numerical results for the various gluon TMDs and discuss several properties of their small-$x$ evolution such as geometric scaling. Finally, section VI is devoted to conclusions and outlook.

\section{Small-x limit of the TMD factorization framework}

\begin{figure}
  \begin{center}
    \includegraphics[width=0.45\textwidth]{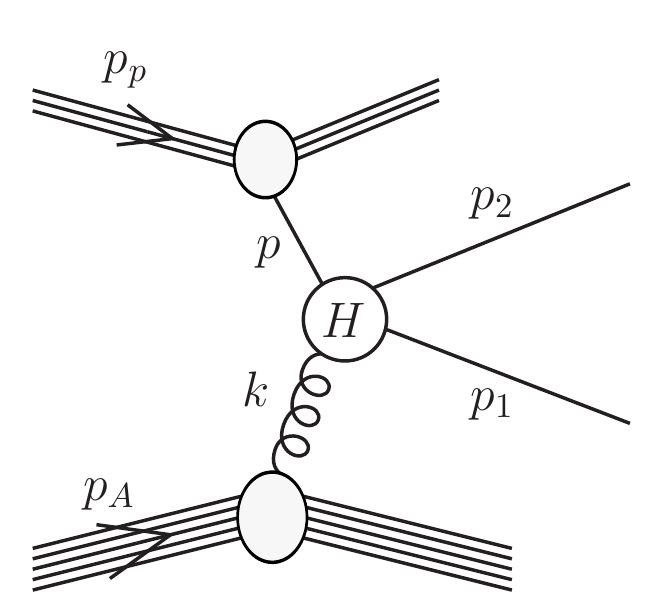}
  \end{center}
  \caption{Inclusive forward di-jet production in p+A collision. The blob $H$ represents
  hard scattering. The solid lines 
  coming out of $H$ represent partons, which can be either quarks or gluons.}
  \label{fig:dijets-pA}
\end{figure}

We consider the process of inclusive di-jet production in the forward region, in collisions of dilute and dense systems
\begin{equation}
  p (p_p) + A (p_A) \to j_1 (p_1) + j_2 (p_2)+ X\ .
\end{equation}
The process is shown schematically in Fig~\ref{fig:dijets-pA}. The four-momenta of the projectile and the target are massless and purely longitudinal.
In terms of the light cone variables, $x^\pm = (x^0\pm x^3)/\sqrt{2}$, they take the simple form
\begin{equation}
  p_p = \sqrt{\frac{s}{2}}(1,0_t,0)\ ,\quad
  p_A = \sqrt{\frac{s}{2}}(0,0_t,1)\ ,
\label{eq:pp-pA-defs}
\end{equation}
where $s$ is the squared center of mass energy of the p+A system.
The energy (or longitudinal momenta) fractions of the incoming parton (either
a quark or gluon) from the projectile, $x_1$, and the gluon from the target, $x_2$, can be expressed in
terms of the rapidities and transverse momenta of the produced jets as
\be
x_1  = \frac{p_1^+ + p_2^+}{p_p^+}   = \frac{1}{\sqrt{s}} \left(|p_{1t}|
e^{y_1}+|p_{2t}| e^{y_2}\right)\ , \quad
x_2  = \frac{p_1^- + p_2^-}{p_A^-}   = \frac{1}{\sqrt{s}} \left(|p_{1t}| e^{-y_1}+|p_{2t}| e^{-y_2}\right)\ ,
\ee
where $p_{1t}$, $p_{2t}$ are transverse Euclidean two-vectors.
By looking at jets produced in the forward direction, we effectively select
those fractions to be $x_1 \sim 1$ and $x_2 \ll 1$.
Since the target A is probed at low $x_2$, the dominant contributions come from
the subprocesses in which the incoming parton on the target side is a gluon
\begin{equation}
  qg  \to  qg\ ,
  \qquad \qquad 
  gg  \to  q\bar q\ ,
  \qquad \qquad 
  gg  \to  gg\ .
\label{eq:3channels}
\end{equation}

Moreover, the large-$x$ partons of the dilute projectile are described in terms of the usual parton distribution functions of collinear factorization $q(x_1,\mu^2)$ and $g(x_1,\mu^2)$, with a scale dependence given by DGLAP evolution equations, while the small-$x$ gluons of the dense target are described by several transverse-momentum-dependent (TMD) distributions, which evolve towards small values of $x_2$ according to non-linear equations. Indeed, besides its longitudinal
component $k^-\!=\!x_2\sqrt{s/2}$, the momentum of the incoming gluon from the target has in general a non-zero transverse component
\be
k_t = p_{1t}+p_{2t}
\label{eq:ktglue}
\ee
which leads to imbalance of transverse momentum of the produced jets: $|k_t|^2 =|p_{1t}|^2 + |p_{2t}|^2 + 2|p_{1t}||p_{2t}| \cos\Delta\phi$.
The Mandelstam variables of the $2\to 2$ process are:
\bea
  \label{eq:mandelstam}
    \hat{s} & = & (p+k)^2 = (p_1 + p_2)^2=\frac{|P_t|^2}{z(1-z)}\,, \\
    \hat{t} & = &(p_2-p)^2 = (p_1 - k)^2=-\frac{|p_{2t}|^2}{1-z}\,, \\
    \hat{u} & = & (p_1-p)^2 = (p_2 - k)^2=-\frac{|p_{1t}|^2}{z}\,,
\eea
with
\begin{equation}
z=\frac{p_1^+}{p_1^+ + p_2^+} \quad\quad \text{and}
\quad\quad P_t=(1-z)p_{1t}-zp_{2t}\ .
\label{eq:zdef}
\end{equation}
They sum up to $\hat{s}+\hat{t}+\hat{u}=-k_t^2$.

\subsection{The TMD factorization formula for forward di-jets}

Just as collinear factorization, the TMD factorization framework is a "leading-twist" framework valid to leading power of the hard scale,
but it can only be established for a subset of hard processes, compared to collinear factorization. In particular, there exists no general
TMD factorization theorem for jet production in hadron-hadron collisions. However, such a factorization can be established in the
asymmetric ``dilute-dense'' situation considered here, where only one of the colliding hadrons is described by a transverse momentum
dependent (TMD) gluon distribution. Again, selecting di-jet systems produced in the forward direction implies $x_1 \sim 1$ and $x_2 \ll 1$,
which in turn allows us to make that assumption.

In this context, the validity domain of the TMD factorization formula is 
\be
|k_t|,Q_s\ll|p_{1t}|,|p_{2t}|\quad\mbox{or}\quad |k_t|,Q_s\ll|P_t|\ .
\ee
This means that the transverse momentum imbalance between the outgoing particles, Eq.~\eqref{eq:ktglue}, must be much smaller than their individual
transverse momenta, which corresponds to the situation of nearly back-to-back di-jets. The jet momenta must also be much bigger than the other momentum scale in the problem, the saturation scale of the dense target $Q_s(x_2)$, and in practice this is always the case. 

The TMD factorization formula reads \cite{Dominguez:2011wm,Kotko:2015ura}:
\begin{equation}
\frac{d\sigma^{pA\rightarrow {\rm dijets}+X}}{dy_1dy_2d^2p_{1t}d^2p_{2t}} =
\frac{\alpha _{s}^{2}}{(x_1x_2s)^{2}} \sum_{a,c,d} x_1 f_{a/p}(x_1,\mu^2) \sum_i H_{ag\to cd}^{(i)}(z,P_t)\ \mathcal{F}_{ag}^{(i)}(x_2,k_t) 
\frac{1}{1+\delta_{cd}}\,,
\label{eq:tmd-main}
\end{equation}
where $\mathcal{F}_{ag}^{(i)}$ denotes several distinct TMD gluon distributions, with different operator definitions.
Each of them is accompanied by its own hard factor $H_{ag\to cd}^{(i)}$. These were calculated in \cite{Dominguez:2011wm} and expressed in terms of the Mandelstam variables \eqref{eq:mandelstam}. Because of the condition $|k_t|\ll|p_{1t}|,|p_{2t}|$, those hard factors are on-shell (i.e. $|k_t|=0$), and the $k_t$ dependence of the cross-section  comes from the gluon distributions $\mathcal{F}_{ag}^{(i)}(x_2,k_t)$ only.

It was shown in \cite{Kotko:2015ura} that  the offshellness of the small-$x$ gluon can be restored in the hard factors
(i.e. $H_{ag\to cd}^{(i)}(P_t)\to H_{ag^*\to cd}^{(i)}(k_t,P_t)$) in order to extend the validity of formula \eqref{eq:tmd-main} to a wider kinematical range: $Q_s\ll|p_{1t}|,|p_{2t}|$ without any condition on the magnitude of $|k_t|$. But in this work, we stick to the strict TMD limit. Explicitly, the three channels read
(in \eqref{eq:tmd-qg} $p_1$ denotes the momentum of the final-state gluon):
\begin{equation}
\frac{d\sigma(pA\to qg X)}{d^2P_{t}d^2k_{t}dy_1dy_2} =\frac{\alpha_s^2}{2C_F}\frac{z(1\!-\!z)}{P_{t}^{\ 4}}
x_1q(x_1,\mu^2)P_{gq}(z)\left\{\left[(1\!-\!z)^2-\frac{z^2}{N_c^2}\right]{\cal F}_{qg}^{(1)}(x_2,k_t)+{\cal F}_{qg}^{(2)}(x_2,k_t)\right\}\ ,
\label{eq:tmd-qg}
\end{equation}
\begin{eqnarray}
\frac{d\sigma(pA\to q\bar q X)}{d^{2}P_{t}d^{2}k_{t}dy_{1}dy_{2}} =\frac{\alpha_s^2}{2C_F}\frac{z(1\!-\!z)}{P_{t}^{\ 4}}
x_1g(x_1,\mu^2)P_{qg}(z)\left\{[(1\!-\!z)^2+z^2]{\cal F}_{gg}^{(1)}(x_2,k_t)+2z(1-z)\mbox{Re}\ {\cal F}_{gg}^{(2)}(x_2,k_t)\right.\nonumber\\\left.-\frac1{N_c^2}{\cal F}_{gg}^{(3)}(x_2,k_t)\right\}\ ,
\label{eq:tmd-qqbar}
\end{eqnarray}
\begin{eqnarray}
\frac{d\sigma(pA\to gg X)}{d^{2}P_{t}d^{2}k_{t}dy_{1}dy_{2}} =\frac{\alpha_s^2}{2C_F}\frac{z(1\!-\!z)}{P_{t}^{\ 4}}
x_1g(x_1,\mu^2)P_{gg}(z)\left\{[(1\!-\!z)^2+z^2]{\cal F}_{gg}^{(1)}(x_2,k_t)+2z(1-z)\mbox{Re}\ {\cal F}_{gg}^{(2)}(x_2,k_t)\right.\nonumber\\\left.
+{\cal F}_{gg}^{(6)}(x_2,k_t)+\frac{1}{N_c^2}\left[{\cal F}_{gg}^{(4)}(x_2,k_t)+{\cal F}_{gg}^{(5)}(x_2,k_t)-2{\cal F}_{gg}^{(3)}(x_2,k_t)\right]\right\}\ ,
\label{eq:tmd-gg}
\end{eqnarray}
with
\begin{equation}
P_{gq} = C_F\ \frac{1+(1\!-\!z)^2}{z}\ ,\quad
P_{qg} = \frac{z^2+(1\!-\!z)^2}2\ , \quad
P_{gg} = 2N_c\left[\frac{z}{1-z} + \frac{1-z}{z} + z(1-z)\right]\ .
\ee

Several gluon distributions $\mathcal{F}_{ag}^{(i)}$, with different operator definition, are involved here. Indeed, a generic unintegrated gluon distribution of the form
\cite{Mulders:2000sh}
\begin{equation}
\mathcal{F}(x_2,k_t) \stackrel{\text{naive}}{=}
2\int \frac{d\xi^+d^2{\boldsymbol\xi}}{(2\pi )^{3}p_A^{-}}e^{ix_2p_A^{-}\xi ^{+}-ik_t\cdot{\boldsymbol\xi}}
\left\langle A|\text{Tr}\left[F^{i-}\left( 0\right)F^{i-}\left(\xi^+,{\boldsymbol\xi}\right)
\right]|A\right\rangle\,,
\label{eq:tmd-naive}
\end{equation}
where $F^{i-}$ are components of the gluon field strength tensor, must be also supplemented with gauge links, in order to render such a bi-local product of field operators gauge invariant \cite{Belitsky:2002sm}. The gauge links are path-ordered exponentials, with the integration path being fixed by the hard part of the process under consideration. In the following, we shall encounter two gauge links $\mathcal{U}^{\left[ +\right] }$ and $\mathcal{U}^{\left[ -\right] }$, as well as loops $\mathcal{U}_0^{\left[\square \right] }=\mathcal{U}^{\left[ +\right] }\mathcal{U}^{\left[ -\right]\dagger}$ and $\mathcal{U}_\xi^{\left[\square \right] }=\mathcal{U}^{\left[ -\right] \dagger}\mathcal{U}^{\left[ +\right]}$. The various gluon distributions needed for the di-jet process are given by \cite{Bomhof:2006dp,Kotko:2015ura}:
\begin{eqnarray}
\mathcal{F}_{qg}^{(1)}(x_2,k_t) &=&2\int \frac{d\xi^+d^2{\boldsymbol\xi}}{(2\pi )^{3}p_A^{-}}e^{ix_2p_A^{-}\xi ^{+}-ik_t\cdot{\boldsymbol\xi}}
\left\langle A\left|\text{Tr}\left[ F^{i-}(\xi)\ \mathcal{U}^{\left[ -\right] \dagger }F^{i-}(0)\ \mathcal{U}^{\left[ +\right] }\right] \right|A\right\rangle\ ,
\label{eq:tmdqg1}\\
\mathcal{F}_{qg}^{(2)}(x_2,k_t) &=&2\int \frac{d\xi^+d^2{\boldsymbol\xi}}{(2\pi )^{3}p_A^{-}}e^{ix_2p_A^{-}\xi ^{+}-ik_t\cdot{\boldsymbol\xi}}\
\frac{1}{N_c}\left\langle A\left|\text{Tr}\left[ F^{i-}(\xi)\ \mathcal{U}^{\left[ +\right] \dagger}F^{i-}(0)\ \mathcal{U}^{\left[ +\right] }\right]
\text{Tr}\left[ \mathcal{U}^{\left[\square \right] }\right]\right|A\right\rangle\ ,
\label{eq:tmdqg2}\\
\mathcal{F}_{gg}^{(1)}(x_2,k_t) &=&2\int \frac{d\xi^+d^2{\boldsymbol\xi}}{(2\pi )^{3}p_A^{-}}e^{ix_2p_A^{-}\xi ^{+}-ik_t\cdot{\boldsymbol\xi}}\
\frac{1}{N_c}\left\langle A\left| \text{Tr}\left[ F^{i-}(\xi)\ \mathcal{U}^{\left[ -\right] \dagger }F^{i-}(0)\ \mathcal{U}^{\left[ +\right] }\right]
\text{Tr}\left[ \mathcal{U}^{\left[\square \right]\dagger }\right]\right|A\right\rangle\ ,
\\
\mathcal{F}_{gg}^{(2)}(x_2,k_t) &=&2\int \frac{d\xi^+d^2{\boldsymbol\xi}}{(2\pi )^{3}p_A^{-}}e^{ix_2p_A^{-}\xi ^{+}-ik_t\cdot{\boldsymbol\xi}}\
\frac{1}{N_c}\left\langle A\left|\text{Tr}\left[ F^{i-}(\xi)\ \mathcal{U}_\xi^{\left[\square\right]\dagger} \right]
\textrm{Tr}\left[ F^{i-}(0)\ \mathcal{U}_0^{\left[ \square\right] }\right]\right|A\right\rangle\ ,
\\
\mathcal{F}_{gg}^{(3)}(x_2,k_t) &=& 2\int \frac{d\xi^+d^2{\boldsymbol\xi}}{(2\pi )^{3}p_A^{-}}e^{ix_2p_A^{-}\xi ^{+}-ik_t\cdot{\boldsymbol\xi}}
\left\langle A\left| \text{Tr}\left[F^{i-}(\xi)\ \mathcal{U}^{\left[+\right] \dagger }F^{i-}(0)\ \mathcal{U}^{\left[ +\right] }\right] \right|A\right\rangle\ ,
\\
\mathcal{F}_{gg}^{(4)}(x_2,k_t) &=& 2\int \frac{d\xi^+d^2{\boldsymbol\xi}}{(2\pi )^{3}p_A^{-}}e^{ix_2p_A^{-}\xi ^{+}-ik_t\cdot{\boldsymbol\xi}}
\left\langle A\left| \text{Tr}\left[F^{i-}(\xi)\ \mathcal{U}^{\left[-\right] \dagger }F^{i-}(0)\ \mathcal{U}^{\left[ -\right] }\right] \right|A\right\rangle\ ,
\\
\mathcal{F}_{gg}^{(5)}(x_2,k_t) &=& 2\int \frac{d\xi^+d^2{\boldsymbol\xi}}{(2\pi )^{3}p_A^{-}}e^{ix_2p_A^{-}\xi ^{+}-ik_t\cdot{\boldsymbol\xi}}
\left\langle A\left| \text{Tr}\left[F^{i-}(\xi)\ \mathcal{U}_\xi^{\left[\square \right] \dagger } \mathcal{U}^{\left[+\right] \dagger }
F^{i-}(0)\ \mathcal{U}_0^{\left[\square \right] } \mathcal{U}^{\left[ +\right] }\right] \right|A\right\rangle\ ,
\\
\mathcal{F}_{gg}^{(6)}(x_2,k_t) &=&2\int \frac{d\xi^+d^2{\boldsymbol\xi}}{(2\pi )^{3}p_A^{-}}e^{ix_2p_A^{-}\xi ^{+}-ik_t\cdot{\boldsymbol\xi}}\
\frac{1}{N^2_c}\left\langle A\left| \text{Tr}\left[ F^{i-}(\xi)\ \mathcal{U}^{\left[ +\right] \dagger }F^{i-}(0)\ \mathcal{U}^{\left[ +\right] }\right]
\text{Tr}\left[ \mathcal{U}^{\left[\square \right] }\right] \text{Tr}\left[ \mathcal{U}^{\left[\square \right] \dagger}\right] \right|A\right\rangle\ .
\label{eq:tmdgg6}
\end{eqnarray}

The gauge links are composed of Wilson lines, their simplest expression is obtained in the $A^+=0$ gauge (but the expressions above are gauge-invariant):
\begin{equation} 
\mathcal{U}^{\left[ \pm\right] }= U(0^+,\pm\infty;{\bf 0})U(\pm\infty,\xi^+;{\boldsymbol\xi})\quad \mbox{with}\quad
U(a,b;{\bf{x}}) = \mathcal{P} \exp \left[ ig \int_a^b dx^+ A_c^-(x^+, {\bf{x}}) t^c \right]\ ,
\end{equation}
where $t^c$ are the generators of the fundamental representation of $SU(N_c)$. The gluon TMDs are normalized such that
$\int d^2k_t\ \mathcal{F}_{ag}^{(i)}(x_2,k_t)=x_2 f_{g/A}(x_2)$, except for $\mathcal{F}_{gg}^{(2)}$ which vanishes when integrated.

\subsection{Taking the small-x limit}

In \eqref{eq:tmd-naive}, the matrix element is calculated for a hadronic/nuclear state with a fixed given momentum $p_A$, normalized such that
$\langle p|p'\rangle=(2\pi )^{3}\ 2p^-\delta(p^- - p'^-)\delta^{(2)}(p_t - p'_t)$. Therefore, using translational invariance, we may write
\be
\int \frac{d\xi^+d^2{\boldsymbol\xi}}{(2\pi )^{3}p_A^{-}}e^{ix_2p_A^{-}\xi ^{+}-ik_t\cdot{\boldsymbol\xi}}\left\langle A|O(0,\xi)|A\right\rangle
=\frac{2}{\langle A|A\rangle}\int\frac{d^3\xi d^3\xi'}{(2\pi )^{3}}e^{ix_2p_A^{-}(\xi ^{+}-\xi ^{'+})-ik_t\cdot({\boldsymbol\xi}-{\boldsymbol\xi}')}
\left\langle A| O(\xi',\xi)|A \right\rangle\ .
\ee
In the small $x_2$ limit, we set $\exp[ix_2p_A^{-}(\xi ^{+}\!-\!\xi ^{'+})]\!=\!1$ and we evaluate the matrix elements as Color Glass Condensate averages, which now contain the $x_2$ dependence:
\be
\frac{\left\langle A| O(\xi',\xi)|A \right\rangle}{\langle A|A\rangle}=\left\langle O(\xi',\xi) \right\rangle_{x_2}\ .
\ee
Then, for instance, we can write for $\mathcal{F}_{qg}^{(1)}$ (see \eqref{eq:tmdqg1}):
\be
\mathcal{F}_{qg}^{(1)}(x_2,k_t) =4\int\frac{d^3x d^3y}{(2\pi)^3}\ e^{-ik_t\cdot({\bf x}-{\bf y})}
\left\langle \text{Tr}\left[ F^{i-}(x)\ \mathcal{U}^{\left[ -\right] \dagger }F^{i-}(y)\ \mathcal{U}^{\left[ +\right] }\right] \right\rangle_{x_2}\ , 
\label{eq:Fqg1}
\ee
and similarly for all the other gluon TMDs \eqref{eq:tmdqg2}-\eqref{eq:tmdgg6}.

More details on their $x_2$ dependence are given below, but first let us simplify further their expressions. From Eq.~\eqref{eq:Fqg1} we have:
\bea 
\mathcal{F}_{qg}^{(1)}(x_2,k_t) &=&4\int\frac{d^3x d^3y}{(2\pi)^3}\ e^{-ik_t\cdot({\bf x}-{\bf y})} 
\left< {\text{Tr}}\, F^{i-} (x) U[x^+, -\infty; {\bf x}] U [-\infty, y^+; {\bf y}] F^{i-} (y) U[y^+, +\infty; {\bf y}]  U[+\infty, x^+; {\bf x}]\right>_{x_2} \nonumber \\
&=& 
 4\int\frac{d^3x d^3y}{(2\pi)^3}\ e^{-ik_t\cdot({\bf x}-{\bf y})}\left< {\text{Tr}}\, U [+\infty, x^+; {\bf x}] F^{i-} (x) U [x^+, -\infty;{\bf x}]
 \right . \nonumber \\&& \left .  \hspace{3.8cm}\times U [-\infty, y^+; {\bf y}]  F^{i-} (y) U [y^+, +\infty; {\bf y}] \right>_{x_2} \, .
\eea
Using the formula for the derivative of the Wilson lines
\be 
\partial_i U_{\bf y} = ig \int_{-\infty}^{\infty} dy^+ U[-\infty, y^+;{\bf y}] F^{i-} (y) U[y^+, +\infty;{\bf y}] \, 
\ee
with $F^{i-}(y) = \partial^i A^-(y)$ in the $A^+=0$ gauge
(otherwise, the other piece of the field strength tensor comes additional transverse gauge links in $\mathcal{U}^{\left[ \pm\right] }$), we obtain
\be
\mathcal{F}_{qg}^{(1)}(x_2,k_t) =\frac{4}{g^2}\int\frac{d^2{\bf x} d^2{\bf y}}{(2\pi)^3}\ e^{-ik_t\cdot({\bf x}-{\bf y})}
\left\langle \text{Tr}\left[ (\partial_i U^\dagger_{\bf x}) (\partial_i U_{\bf y})\right] \right\rangle_{x_2}=
\frac{N_c k_t^2}{2\pi^2\alpha_s}\int\frac{d^2{\bf x} d^2{\bf y}}{(2\pi)^2}\ e^{-ik_t\cdot({\bf x}-{\bf y})}\frac1{N_c}
\left\langle \text{Tr}\left[U_{\bf y} U^\dagger_{\bf x}\right] \right\rangle_{x_2}\ ,
\label{eq:fqg1}
\ee
where
\be
U_{\bf x}\equiv U(-\infty,+\infty;{\bf x})=\mathcal{P} \exp \left[ ig \int_{-\infty}^{\infty} dx^+ A_a^-(x^+, {\bf{x}}) t^a \right]\ .
\ee
Due to its simple Wilson line structure, $\mathcal{F}_{qg}^{(1)}$ has been dubbed the "dipole" gluon distribution.

To give a second example which leads to a more complicated Wilson line structure, let us also simplify the so-called Weizs\"acker-Williams gluon distribution $\mathcal{F}_{gg}^{(3)}$:
\bea 
\mathcal{F}_{gg}^{(3)}(x_2,k_t) &=&4\int\frac{d^3x d^3y}{(2\pi)^3}\ e^{-ik_t\cdot({\bf x}-{\bf y})} 
\left< {\text{Tr}}\, F^{i-} (x) U[x^+, +\infty; {\bf x}] U[+\infty, y^+; {\bf y}] F^{i-} (y) U[y^+, +\infty;{\bf y}] U[+\infty, x^+;{\bf x}]\right>_{x_2} \nonumber \\
&=& 
 4\int\frac{d^3x d^3y}{(2\pi)^3}\ e^{-ik_t\cdot({\bf x}-{\bf y})}\left< {\text{Tr}}\, U [-\infty, x^+; {\bf x}]  F^{i-} (x) U [x^+, +\infty;{\bf x}]U^\dagger [-\infty, +\infty; {\bf y}] 
  \right . \nonumber \\ && \left .  \hspace{3.8cm}\times \, U [-\infty, y^+; {\bf y}] F^{i-} (y) U [y^+, +\infty;{\bf y}]U^\dagger [-\infty, +\infty; {\bf x}] \right>_{x_2} \nonumber \\
&=& 
- \frac{4}{g^2}\int\frac{d^2{\bf x} d^2{\bf y}}{(2\pi)^3}\ e^{-ik_t\cdot({\bf x}-{\bf y})}
\left\langle \text{Tr}\left[ (\partial_i U_{\bf x}) U^\dagger_{\bf y} (\partial_i U_{\bf y}) U^\dagger_{\bf x}  \right] \right\rangle_{x_2} \, .
\label{eq:fgg3}
\eea

Following the same lines, we obtain for the other gluon TMDs:
\begin{eqnarray}
\mathcal{F}_{qg}^{(2)}(x_2,k_t) &=& -\frac{4}{g^2}\int\frac{d^2{\bf x} d^2{\bf y}}{(2\pi)^3}\ e^{-ik_t\cdot({\bf x}-{\bf y})}
\frac1{N_c}
\left\langle \text{Tr}\left[ (\partial_i U_{\bf x}) U^\dagger_{\bf y} (\partial_i U_{\bf y}) U^\dagger_{\bf x}  \right] \text{Tr}\left[ U_{\bf y} U^\dagger_{\bf x}\right] \right\rangle_{x_2}\ ,
\label{eq:fqg2}\\
\mathcal{F}_{gg}^{(1)}(x_2,k_t) &=& \frac{4}{g^2}\int\frac{d^2{\bf x} d^2{\bf y}}{(2\pi)^3}\ e^{-ik_t\cdot({\bf x}-{\bf y})}
\frac1{N_c}
\left\langle \text{Tr}\left[  (\partial_i U_{\bf y}) (\partial_i U^\dagger_{\bf x}) \right] \text{Tr}\left[ U_{\bf x} U^\dagger_{\bf y}\right] \right\rangle_{x_2}\ ,
\label{eq:fgg1}\\
\mathcal{F}_{gg}^{(2)}(x_2,k_t) &=& -\frac{4}{g^2}\int\frac{d^2{\bf x} d^2{\bf y}}{(2\pi)^3}\ e^{-ik_t\cdot({\bf x}-{\bf y})}
\frac1{N_c}
\left\langle \text{Tr}\left[ (\partial_i U_{\bf x}) U^\dagger_{\bf y}\right] \text{Tr}\left[ (\partial_i U_{\bf y}) U^\dagger_{\bf x}  \right] \right\rangle_{x_2}\ ,
\label{eq:fgg2}\\
\mathcal{F}_{gg}^{(4)}(x_2,k_t) &=&-\frac{4}{g^2}\int\frac{d^2{\bf x} d^2{\bf y}}{(2\pi)^3}\ e^{-ik_t\cdot({\bf x}-{\bf y})}
\left\langle \text{Tr}\left[ (\partial_i U_{\bf x}) U^\dagger_{\bf x} (\partial_i U_{\bf y}) U^\dagger_{\bf y}  \right] \right\rangle_{x_2}\ ,
\label{eq:fgg4}\\
\mathcal{F}_{gg}^{(5)}(x_2,k_t) &=& -\frac{4}{g^2}\int\frac{d^2{\bf x} d^2{\bf y}}{(2\pi)^3}\ e^{-ik_t\cdot({\bf x}-{\bf y})}
\left\langle \text{Tr}\left[ (\partial_i U_{\bf x}) U^\dagger_{\bf y} U_{\bf x} U^\dagger_{\bf y} (\partial_i U_{\bf y}) U^\dagger_{\bf x} U_{\bf y} U^\dagger_{\bf x} \right]
\right\rangle_{x_2}\ ,
\label{eq:fgg5}\\
\mathcal{F}_{gg}^{(6)}(x_2,k_t) &=& -\frac{4}{g^2}\int\frac{d^2{\bf x} d^2{\bf y}}{(2\pi)^3}\ e^{-ik_t\cdot({\bf x}-{\bf y})}
\frac1{N_c^2}
\left\langle \text{Tr}\left[ (\partial_i U_{\bf x}) U^\dagger_{\bf y} (\partial_i U_{\bf y}) U^\dagger_{\bf x}  \right] \text{Tr}\left[ U_{\bf x} U^\dagger_{\bf y}\right]
\text{Tr}\left[ U_{\bf y} U^\dagger_{\bf x}\right] \right\rangle_{x_2}\ .
\label{eq:fgg6}
\end{eqnarray}

The CGC averages $\langle\ \cdot\ \rangle_{x_2}$ are averages over
color field configurations in the dense target. They may be written
\be
\left\langle O \right\rangle_{x_2}=\int DA^- |\phi_{x_2}[A^-]|^2 O[A^-]\ ,
\ee
where $|\phi_{x_2}[A^-]|^2$ represents the probability of a given field configuration. The CGC wavefunction $\phi_{x_2}[A^-]$ effectively describes, in terms of strong classical fields, the dense parton content of a hadronic/nuclear wave function, at small longitudinal momentum fraction $x_2$. In the leading-logarithmic approximation, the evolution of $|\phi_{x_2}[A^-]|^2$ with decreasing $x_2$ is obtained from the JIMWLK equation:
\bea
\frac{d}{d\log(1/x_2)} |\phi_{x_2}[A^-]|^2 &=&
\int
\frac{d^2{\bf x}}{2\pi}\frac{d^2{\bf y}}{2\pi}\frac{d^2{\bf z}}{2\pi}\frac{({\bf x}\!-\!{\bf z})\cdot({\bf y}\!-\!{\bf z})}{({\bf x}\!-\!{\bf z})^2({\bf z}\!-\!{\bf y})^2}
\frac{\delta}{\delta A^-_c({\bf x})}\left[1+V^\dagger_{\bf x}V_{\bf y}-V^\dagger_{\bf x}V_{\bf z}-V^\dagger_{\bf z}V_{\bf y}\right]^{cd}
\frac{\delta}{\delta A^-_d({\bf y})} |\phi_{x_2}[A^-]|^2\hspace{0.5cm}\\
&=& H_{JIMWLK} |\phi_{x_2}[A^-]|^2\ ,
\eea
where the functional derivatives $\delta/\delta A^-_c({\bf x})$ act at the largest value of $x^+$:
\be
\frac{\delta}{\delta A^-_c({\bf x})}\equiv \lim_{x^+\to\infty}\frac{\delta}{\delta A^-_c(x^+,{\bf x})}\ ,
\ee
and where
\be 
V_{\bf x} = \mathcal{P} \exp \left[ ig \int_{-\infty}^{\infty}  dx^+ A_a^-(x^+, {\bf{x}}) T^a
\right]
\ee
with $T^a$ denoting the generators of the adjoint representation of $SU(N_c)$. After integrating by parts, the evolution of any CGC average may be written:
\be
\frac{d}{d\log(1/x_2)} \left\langle O \right\rangle_{x_2}  = \left\langle H_{JIMWLK}\ O \right\rangle_{x_2}\ .
\ee

We note that recently, more general evolution equations have been derived for the gluon TMDs $\mathcal{F}_{gg}^{(3)}$ \cite{Balitsky:2015qba,Zhou:2016tfe} and
$\mathcal{F}_{gg}^{(4)}$ \cite{Balitsky:2016dgz}. These equations contain JIMWLK evolution in the small-$x$ limit (or
Balitsky-Fadin-Kuraev-Lipatov (BFKL) evolution \cite{Lipatov:1976zz,Kuraev:1976ge,Balitsky:1978ic} in the linear approximation), and in addition for generic values of $x$, the hard scale dependence, which at small $x$ boils down to Sudakov factors \cite{Mueller:2012uf,Mueller:2013wwa}. Presumably, similar equations could also be obtained for the other gluon TMDs. It is not known how to solve them however, and in this work we stick to the small-$x$ JIMWLK evolution.

In section \ref{sec:lattice} we shall evaluate numerically all the gluon TMDs given above, using a lattice calculation to solve the JIMWLK equation. But first, we will demonstrate that the small-$x$ limit of the TMD factorization formula \eqref{eq:tmd-main} with those gluon TMDs can also be obtained directly in the CGC framework.

\section{Leading power of the CGC framework}

In this section, our starting point is CGC formalism for double-inclusive particle production in dilute-dense collisions.
We shall extract the leading power in $1/P_t^2$ and show that the result coincides with the small-$x$ limit of TMD factorization formula for
forward di-jets \eqref{eq:tmd-main}, namely Eq.~\eqref{eq:tmd-qg}-\eqref{eq:tmd-gg} with the gluon TMDs given by \eqref{eq:fqg1}, \eqref{eq:fgg3} and \eqref{eq:fqg2}-\eqref{eq:fgg6}. In the large-$N_c$ limit, this has already been demonstrated in \cite{Dominguez:2011wm} for all three channels (see \eqref{eq:3channels}), in \cite{Akcakaya:2012si} for the $g\to q\bar q$ case and in \cite{Iancu:2013dta} for the $g\to gg$ subprocess.

In the present work, we show the equivalence between the CGC and TMD framework, in their overlapping domain of availability, while keeping $N_c$ finite. We start with the quark initiated channel, for which we shall explain the derivation in details, and then we deal with the gluon initiated channels.

\subsection{The quark initiated channel $q\to qg$}

\begin{figure}
  \begin{center}
    \includegraphics[width=0.80\textwidth]{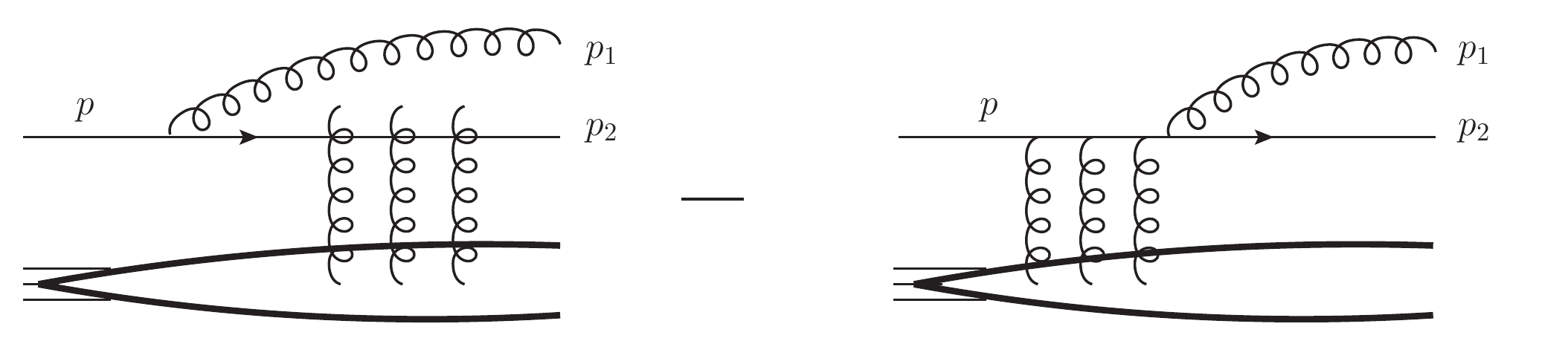}
  \end{center}
  \caption{Amplitude for quark-gluon production in the CGC formalism. Left: the gluon is radiated before the interaction with the target. Right: the gluon is radiated after the interaction with the target. The two terms have a relative minus sign.}
  \label{fig:amplitudeCGC}
\end{figure}

The amplitude for quark-gluon production is schematically presented in Fig.~\ref{fig:amplitudeCGC} as in Ref.~\cite{Marquet:2007vb}.
In the CGC formalism, the scattering of the partons from the dilute projectile with the dense target is described by Wilson lines that resum
multi-gluon exchanges; fundamental Wilson lines for quarks and adjoint Wilson lines for gluons. As a result, the cross section involves
multipoint correlators of Wilson lines. In particular, the square of the amplitude from Fig.~\ref{fig:amplitudeCGC} contains four terms:
a correlator of four Wilson lines, $S^{(4)}$, corresponding to interactions happening after the emission of the gluon, both in the amplitude
and the complex conjugate, then a correlator of two Wilson lines, $S^{(2)}$,  representing the case when interactions with the target take
place before the radiation of the gluon in both amplitude and complex conjugate, and two correlators of three Wilson lines, $S^{(3)}$, for the cross terms.

Denoting, as in the previous section, $p_1$ the momentum of the outgoing gluon and $p_2$ the momentum of the
outgoing quark, the cross-section reads \cite{Marquet:2007vb}:
\bea
\frac{d\sigma(pA\to qgX)}{d^2p_{1t} d^2p_{2t}dy_1 dy_2} = \alpha_s C_F (1-z) p_1^+ x_1 q(x_1,\mu^2)
\int\frac{d^2{\bf u}}{(2\pi)^2}\frac{d^2{\bf u'}}{(2\pi)^2} e^{i P_t \cdot ({\bf u'}-{\bf u})}
\sum_{\lambda\alpha\beta} \phi^{\lambda^*}_{\alpha\beta}(p,p_1^+,{\bf u'}) \phi^{\lambda}_{\alpha\beta}(p,p_1^+,{\bf u})\hspace{1cm}
\nonumber\\\int\frac{d^2{\bf v}}{(2\pi)^2}\frac{d^2{\bf v'}}{(2\pi)^2} e^{i k_t \cdot ({\bf v'}-{\bf v})}
\left\{S^{(4)}_{qg\bar{q}g}\left({\bf b},{\bf x},{\bf b'},{\bf x'};x_2\right)-S^{(3)}_{qg\bar{q}}\left({\bf b},{\bf x},{\bf v'};x_2\right)
-S^{(3)}_{qg\bar{q}}\left({\bf v},{\bf x'},{\bf b'},x_2\right)+S^{(2)}_{q\bar{q}}\left({\bf v},{\bf v'};x_2\right)\right\}\ ,
\label{eq:cgc-qg}
\eea
where
\be
{\bf b}={\bf v}-z{\bf u}\quad\mbox{and}\quad{\bf b'}={\bf v'}-z{\bf u'}
\ee
denote the transverse positions of the final-state quark in the amplitude and the conjugate amplitude, respectively, and
\be
{\bf x}={\bf v}+(1\!-\!z){\bf u}\quad\mbox{and}\quad{\bf x'}={\bf v'}+(1\!-\!z){\bf u'}
\ee
denote the transverse positions of the final-state gluon in the amplitude and the conjugate amplitude, respectively. ${\bf{u'}}-{\bf{u}}$ is conjugate to $P_t=(1-z)p_{1t}-zp_{2t}$, and ${\bf{v'}}-{\bf{v}}$ is conjugate to the total transverse momentum of the produced particles $k_t=p_{1t}+p_{2t}$.

The $S^{(i)}$ Wilson line correlators are given by:
\bea
S^{(4)}_{qg\bar{q}g}({\bf{b}},{\bf{x}},{\bf{b'}},{\bf{x'}};x_2)&=&\frac{1}{C_F N_c}\left<\text {Tr}\left(U_{\bf b}U^\dagger_{\bf b'}t^dt^c\right)
\left[V_{\bf x}V^\dagger_{\bf x'}\right]^{cd}\right>_{x_2}\ ,\\
S^{(3)}_{qg\bar{q}}({\bf{b}},{\bf{x}},{\bf{b'}};x_2)&=&\frac{1}{C_F N_c}\left<\text {Tr}\left(U^\dagger_{\bf b'}t^cU_{\bf b}t^d\right)V^{cd}_{\bf x}\right>_{x_2}\ ,\\
S^{(2)}_{q\bar{q}}({\bf{b}},{\bf{b'}};x_2)&=&\frac{1}{N_c}\left<\text {Tr} \left(U_{\bf b}U^\dagger_{\bf b'}\right)\right>_{x_2}\ ,
\eea
and the functions $\phi^{\lambda}_{\alpha\beta}$ denote the $q\to qg$ splitting wave functions. In the limit of massless quarks, the wave function overlap is simply given by:
\be 
\sum_{\lambda\alpha\beta} \phi^{\lambda^*}_{\alpha\beta}(p,p^+_1,{\bf{u'}}) \phi^{\lambda}_{\alpha\beta}(p,p^+_1,{\bf{u}}) =
\frac{8\pi^2}{p^+_1} \frac{{\bf{u}} \cdot {\bf{u'}}}{|{\bf{u}}|^2  |{\bf{u'}}|^2} [1 + (1-z)^2]\ .
\ee

\subsection{Extracting the leading power}

In the $|k_t|,Q_s\ll|P_t|$ limit, the integrals in \eqref{eq:cgc-qg} are controlled by configurations where $|{\bf u}|$ and $|{\bf u'}|$ are small compared to the other transverse-size variables, and the leading $1/P^2_t$ power of this expression can be extracted by expanding around ${\bf b}={\bf x}={\bf v}$ and ${\bf b'}={\bf x'}={\bf v'}$. To do this, let us first rewrite all the Wilson line correlators in terms of fundamental Wilson lines only:
\bea
S^{(4)}_{qg\bar{q}g}({\bf{b}},{\bf{x}},{\bf{b'}},{\bf{x'}})&=&
\frac{N_c}{2C_F}\left < Q({\bf b},{\bf b'},{\bf x'},{\bf x})D({\bf x},{\bf x'}) -\frac{1}{N_c^2} D({\bf b},{\bf b'})\right>_{x_2}\ ,\\
S^{(3)}_{qg\bar{q}} ({\bf{b}},{\bf{x}},{\bf{b'}}) &=& \frac{N_c}{2C_F} \left < D({\bf b},{\bf x})D({\bf x},{\bf b'}) -\frac{1}{N_c^2}D({\bf b},{\bf b'})\right>_{x_2}\ ,\\
S^{(2)}_{q\bar{q}}({\bf{b}},{\bf{b'}})&=&\Big<D({\bf b},{\bf b'})\Big>_{x_2}\ ,
\eea
where
\be
D({\bf x},{\bf y})=\frac{1}{N_c}{\text{Tr}}\left(U_{\bf x} U^\dagger_{\bf y}\right)\quad\mbox{and}\quad
Q({\bf x},{\bf y},{\bf v},{\bf w})=\frac{1}{N_c}{\text {Tr}}\left(U_{\bf x} U^\dagger_{\bf y} U_{\bf v} U^\dagger_{\bf w}\right)\ .
\ee
Then, the combination inside the brackets $\Big\{ . \Big\}$ in Eq.~\eqref{eq:cgc-qg} can be rewritten:
\bea
\frac{N_c}{2C_F}\Big<
Q[{\bf v}\!-\!z{\bf u},{\bf v'}\!-\!z{\bf u'},{\bf v'}\!+\!(1\!-\!z){\bf u'},{\bf v}\!+\!(1\!-\!z){\bf u}]D[{\bf v}\!+\!(1\!-\!z){\bf u},{\bf v'}\!+\!(1\!-\!z){\bf u'}]+D[{\bf v},{\bf v'}]\nonumber\\
-D[{\bf v}\!-\!z{\bf u},{\bf v}\!+\!(1\!-\!z){\bf u}]D[{\bf v}\!+\!(1\!-\!z){\bf u},{\bf v'}]
-D[{\bf v},{\bf v'}\!+\!(1\!-\!z){\bf u'}]D[{\bf v'}\!+\!(1\!-\!z){\bf u'},{\bf v'}\!-\!z{\bf u'}]\Big>_{x_2}\nonumber\\
-\frac{1}{2C_F N_c}\Big<D[{\bf v}\!-\!z{\bf u},{\bf v'}\!-\!z{\bf u'}]-D[{\bf v}\!-\!z{\bf u},{\bf v'}]-D[{\bf v},{\bf v'}\!-\!z{\bf u'}]+D[{\bf v},{\bf v'}]\Big>_{x_2}
\eea

This expression vanishes if either {\bf u} or {\bf u'} is set to zero. Therefore, the first non-zero term in its expansion is the one that contains both one power of ${\bf u}$ and one power of ${\bf u'}$:
\be
\frac{N_c u^i u'^j}{2C_F}  [(1\!-\!z)\partial^i_v-z\partial^i_x] [(1\!-\!z)\partial^j_{v'}-z\partial^j_y]
\Big<Q({\bf x},{\bf y},{\bf v'},{\bf v})D({\bf v},{\bf v'})\Big>_{x_2}\Big|_{\substack{{\bf x}={\bf v}\\{\bf y}={\bf v'}}}
-\frac{z^2 u^i u'^j}{2C_FN_c} \partial^i_v\partial^j_{v'}\Big<D({\bf v},{\bf v'})\Big>_{x_2}\ .
\label{eq:lt-qg}
\ee
With the help of the identities $U_{\bf x} \partial U^\dagger_{\bf x}=-(\partial U_{\bf x}) U^\dagger_{\bf x}$ and $\text{Tr}\left[ U_{\bf x} \partial U^\dagger_{\bf x}\right] =0$, Eq.~\eqref{eq:lt-qg} can be rewritten
\be
\frac{N_c u^i u'^j}{2C_F} \left[(1-z)^2-\frac{z^2}{N_c^2}\right]\partial^i_v\partial^j_{v'}\Big<D({\bf v},{\bf v'})\Big>_{x_2}
-\frac{N_c u^i u'^j}{2C_F} \Big<D({\bf v},{\bf v'})\partial^i_x\partial^j_{v'}Q({\bf x},{\bf y},{\bf v'},{\bf v}) \Big>_{x_2}\Big|_{\substack{{\bf x}={\bf v}\\{\bf y}={\bf v'}}}\ .
\ee

Then using
\be
\int\frac{d^2{\bf u}}{(2\pi)^2}\frac{d^2{\bf u'}}{(2\pi)^2} e^{i P_t \cdot ({\bf u'}-{\bf u})}\frac{{\bf u} \cdot {\bf u'}}{|{\bf u}|^2  |{\bf u'}|^2}u^i u'^j =
\frac{\delta^{ij}}{4\pi^2 P_t^4}
\ee
and putting all the pieces together, we finally recover
\bea
\frac{d\sigma(pA\to qgX)}{d^2P_{t}d^2k_{t}dy_1dy_2} &=& \frac{N_c \alpha_s}{C_F} \frac{z(1-z)}{P_t^4} x_1q(x_1,\mu^2)P_{gq}(z)
\int\frac{d^2{\bf v}}{(2\pi)^2}\frac{d^2{\bf v'}}{(2\pi)^2} e^{i k_t \cdot ({\bf v'}-{\bf v})}\nonumber\\
&&\left\{\left[(1-z)^2-\frac{z^2}{N_c^2}\right]\partial^i_v\partial^i_{v'}\Big<D({\bf v},{\bf v'})\Big>_{x_2}
-\Big<D({\bf v},{\bf v'})\partial^i_x\partial^i_{v'}Q({\bf x},{\bf y},{\bf v'},{\bf v}) \Big>_{x_2}\Big|_{\substack{{\bf x}={\bf v}\\{\bf y}={\bf v'}}}
\right\}\ .
\eea
which coincides with the small-$x$ limit of the TMD formula \eqref{eq:tmd-qg}.

\subsection{The gluon initiated channels $g\to q\bar q$ and $g\to gg$}

For the gluon-initiated channels, the derivation is performed along the same lines as in the previous subsection, and we shall outline the main steps here.
Starting with the $g\to q\bar q$ channel, the cross-section reads (denoting $p_1$ and $p_2$ the momenta of the outgoing quark and antiquark, respectively) \cite{Dominguez:2011wm}:
\bea
\frac{d\sigma(pA\to q\bar q X)}{d^2p_{1t} d^2p_{2t}dy_1dy_2} = \frac{\alpha_s}{2}(1-z) p_1^+ x_1 g(x_1,\mu^2)
\int\frac{d^2{\bf u}}{(2\pi)^2}\frac{d^2{\bf u'}}{(2\pi)^2} e^{i P_t \cdot ({\bf u'}-{\bf u})}
\sum_{\lambda\alpha\beta} \varphi^{\lambda^*}_{\alpha\beta}(p,p_1^+,{\bf u'}) \varphi^{\lambda}_{\alpha\beta}(p,p_1^+,{\bf u})\hspace{1cm}
\nonumber\\\int\frac{d^2{\bf v}}{(2\pi)^2}\frac{d^2{\bf v'}}{(2\pi)^2} e^{i k_t \cdot ({\bf v'}-{\bf v})}
\left\{S^{(4)}_{q\bar{q}\bar{q}q}\left({\bf x},{\bf b},{\bf x'},{\bf b'};x_2\right)-S^{(3)}_{qg\bar{q}}\left({\bf x},{\bf v'},{\bf b};x_2\right)
-S^{(3)}_{qg\bar{q}}\left({\bf b'},{\bf v},{\bf x'},x_2\right)+S^{(2)}_{gg}\left({\bf v},{\bf v'};x_2\right)\right\}\ ,
\label{eq:cgc-qqbar}
\eea
where
\be
S^{(4)}_{q\bar{q}\bar{q}q}({\bf{x}},{\bf{b}},{\bf{x'}},{\bf{b'}};x_2)
=\frac{1}{C_F N_c}\left<{\text {Tr}} \left(U^\dagger_{\bf b} t^c U_{\bf x} U^\dagger_{\bf x'} t^c U_{\bf b'}\right)\right>_{x_2}\quad\mbox{and}\quad
S^{(2)}_{gg}({\bf{b}},{\bf{b'}};x_2)=\frac{1}{N^2_c-1}\left<{\text {Tr}} \left(V_{\bf b} V^\dagger_{\bf b'} \right)\right>_{x_2}\ ,
\ee
and with the product of $g\to q\bar q$ splitting wave functions in the massless quarks limit given by: 
\be 
\sum_{\lambda\alpha\beta} \varphi^{\lambda^*}_{\alpha\beta}(p,p^+_1,{\bf{u'}}) \varphi^{\lambda}_{\alpha\beta}(p,p^+_1,{\bf{u}}) =
\frac{8\pi^2}{p^+} \frac{{\bf{u}} \cdot {\bf{u'}}}{|{\bf{u}}|^2  |{\bf{u'}}|^2} [z^2 + (1-z)^2]\ .
\ee

In terms of fundamental Wilson lines, we have:
\bea
S^{(4)}_{q\bar{q}\bar{q}q}({\bf{x}},{\bf{b}},{\bf{x'}},{\bf{b'}})&=&
\frac{N_c}{2C_F}\left < D({\bf x},{\bf x'})D({\bf b'},{\bf b}) -\frac{1}{N_c^2} Q({\bf x},{\bf x'},{\bf b'},{\bf b}) \right>_{x_2}\ ,\\
S^{(2)}_{gg}({\bf{b}},{\bf{b'}})&=&\frac{N_c}{2C_F}\left < D({\bf b},{\bf b'})D({\bf b'},{\bf b}) -\frac{1}{N_c^2}\right>_{x_2}\ ,
\eea
and therefore the combination inside the brackets $\Big\{ . \Big\}$ in Eq.~\eqref{eq:cgc-qqbar} can be rewritten:
\bea
\frac{N_c}{2C_F}\Big<D[{\bf v}\!+\!(1\!-\!z){\bf u},{\bf v'}\!+\!(1\!-\!z){\bf u'}]D[{\bf v'}\!-\!z{\bf u'},{\bf v}\!-\!z{\bf u}]+D[{\bf v},{\bf v'}]D[{\bf v'},{\bf v}]\nonumber\\
-D[{\bf v}\!+\!(1\!-\!z){\bf u},{\bf v'}]D[{\bf v'},{\bf v}\!-\!z{\bf u}]-D[{\bf v'}\!-\!z{\bf u'},{\bf v}]D[{\bf v},{\bf v'}\!+\!(1\!-\!z){\bf u'}]\Big>_{x_2}\\
-\frac{1}{2C_F N_c}\Big<1+ Q[{\bf v}\!+\!(1\!-\!z){\bf u},{\bf v'}\!+\!(1\!-\!z){\bf u'},{\bf v'}\!-\!z{\bf u'},{\bf v}\!-\!z{\bf u}]\nonumber\\
-D[{\bf v}\!+\!(1\!-\!z){\bf u},{\bf v}\!-\!z{\bf u}]-D[{\bf v'}\!-\!z{\bf u'},{\bf v'}\!+\!(1\!-\!z){\bf u'}]\Big>_{x_2}\ .
\label{eq:lt-qqbar}
\eea
As before, this expression vanishes if either {\bf u} or {\bf u'} is set to zero, and the first non trivial order in the expansion reads:
\bea
\frac{N_c u^i u'^j}{2C_F}  [(1\!-\!z)\partial^i_v-z\partial^i_x] [(1\!-\!z)\partial^j_{v'}-z\partial^j_y]
\Big<D({\bf v},{\bf v'})D({\bf y},{\bf x})-\frac{1}{N_c^2}Q({\bf v},{\bf v'},{\bf y},{\bf x})\Big>_{x_2}\Big|_{\substack{{\bf x}={\bf v}\\{\bf y}={\bf v'}}}
\\=\frac{N_c u^i u'^j}{2C_F}\left[ z^2 \Big<D({\bf v},{\bf v'})\partial^i_v\partial^j_{v'}D({\bf v'},{\bf v}) \Big>_{x_2}
+ (1-z)^2 \Big<D({\bf v},{\bf v'})\partial^i_v\partial^j_{v'}D({\bf v'},{\bf v}) \Big>_{x_2}^*\right.\\\left.
- 2z(1-z)\mbox{Re} \Big<\left[\partial^i_vD({\bf v},{\bf v'})\right]\partial^j_{v'}D({\bf v'},{\bf v}) \Big>_{x_2}
+\frac{1}{N^2_c} \Big<\partial^i_v\partial^j_{y}Q({\bf v},{\bf v'},{\bf y},{\bf x})\Big>_{x_2}\Big|_{\substack{{\bf x}={\bf v}\\{\bf y}={\bf v'}}}\right]\ .\\
\eea
Putting everything back together, the small-$x$ limit of the TMD formula \eqref{eq:tmd-qqbar} is recovered.

Now on to the $g\to gg$ channel. The cross-section reads
\bea
\frac{d\sigma(pA\to ggX)}{dy_1 dy_2 d^2p_{1t} d^2p_{2t}} = \alpha_s N_c (1-z) p_1^+ x_1 g(x_1,\mu^2)
\int\frac{d^2{\bf u}}{(2\pi)^2}\frac{d^2{\bf u'}}{(2\pi)^2} e^{i P_t \cdot ({\bf u'}-{\bf u})}
\sum_{\lambda\alpha\beta} \psi^{\lambda^*}_{\alpha\beta}(p,p_1^+,{\bf u'}) \psi^{\lambda}_{\alpha\beta}(p,p_1^+,{\bf u})\hspace{1cm}
\nonumber\\\int\frac{d^2{\bf v}}{(2\pi)^2}\frac{d^2{\bf v'}}{(2\pi)^2} e^{i k_t \cdot ({\bf v'}-{\bf v})}
\left\{S^{(4)}_{gggg}\left({\bf b},{\bf x},{\bf b'},{\bf x'};x_2\right)-S^{(3)}_{ggg}\left({\bf b},{\bf x},{\bf v'};x_2\right)
-S^{(3)}_{ggg}\left({\bf v},{\bf x'},{\bf b'},x_2\right)+S^{(2)}_{gg}\left({\bf v},{\bf v'};x_2\right)\right\}\ ,
\label{eq:cgc-gg}
\eea
where
\bea
S^{(4)}_{gggg}({\bf{b}},{\bf{x}},{\bf{b'}},{\bf{x'}};x_2)
&=&\frac{1}{N_c(N_c^2-1)}f^{abc}f^{ade}\left< V_{\bf b}^{bf}V_{\bf x}^{cg}V_{\bf b'}^{df}V_{\bf x'}^{eg} \right>_{x_2}\ ,\\
S^{(3)}_{ggg}({\bf b},{\bf x},{\bf b'};x_2)&=&\frac{1}{N_c(N_c^2-1)} f^{abc}f^{def} \left< V_{\bf b}^{ad}V_{\bf x}^{be}V_{\bf b'}^{cf} \right>_{x_2}\ ,
\eea
and with the product of $g\to gg$ splitting wave functions given by: 
\be 
\sum_{\lambda\alpha\beta} \psi^{\lambda^*}_{\alpha\beta}(p,p^+_1,{\bf{u'}}) \psi^{\lambda}_{\alpha\beta}(p,p^+_1,{\bf{u}}) =
\frac{16\pi^2}{p^+} \frac{{\bf{u}} \cdot {\bf{u'}}}{|{\bf{u}}|^2  |{\bf{u'}}|^2} \left[\frac{z}{1-z}+\frac{1-z}{z}+z(1-z)\right]\ .
\ee

Due to the identities $S^{(4)}_{gggg}({\bf{b}},{\bf{x}},{\bf{b'}},{\bf{b'}})=S^{(3)}_{ggg}({\bf b},{\bf x},{\bf b'})$ and $S^{(3)}_{ggg}({\bf b},{\bf x},{\bf b'})=S^{(2)}_{gg}({\bf b},{\bf b'})$, the expression in the brackets $\Big\{ . \Big\}$ once again vanishes if either {\bf u} or {\bf u'} is set to zero, and the leading $1/P_t^2$ power can be extracted from:
\bea
u^i u'^j [(1\!-\!z)\partial^i_v-z\partial^i_x] [(1\!-\!z)\partial^j_{v'}-z\partial^j_y] S^{(4)}_{gggg}({\bf{x}},{\bf{v}},{\bf{y}},{\bf{v'}})
\Big|_{\substack{{\bf x}={\bf v}\\{\bf y}={\bf v'}}}\ .
\label{eq:lt-gg}
\eea

Writing the $S^{(4)}$ correlator in terms of fundamental Wilson lines only, we obtain:
\be
S^{(4)}_{gggg}({\bf{b}},{\bf{x}},{\bf{b'}},{\bf{x'}})=
\frac{N_c}{4C_F}\Big< Q({\bf b},{\bf b'},{\bf x'},{\bf x})D({\bf b'},{\bf b})D({\bf x},{\bf x'})
-\frac{1}{N^2_c}O({\bf b},{\bf b'},{\bf x'},{\bf x},{\bf b'},{\bf b},{\bf x},{\bf x'})\Big>_{x_2}
+ \mbox{c.c.} 
\ee
where
\be
O({\bf x},{\bf y},{\bf v},{\bf w},{\bf x'},{\bf y'},{\bf v'},{\bf w'})=
\frac{1}{N_c}{\text {Tr}}\left(U_{\bf x} U^\dagger_{\bf y} U_{\bf v} U^\dagger_{\bf w}U_{\bf x'} U^\dagger_{\bf y'} U_{\bf v'} U^\dagger_{\bf w'}\right)\ ,
\ee
and \eqref{eq:lt-gg} turns into:
\bea
\frac{N_c u^i u'^j}{2C_F}\left\{
[z^2+(1-z)^2] \mbox{Re}\ \Big<D({\bf v},{\bf v'})\partial^i_v\partial^j_{v'}D({\bf v'},{\bf v}) \Big>_{x_2}
- 2z(1-z)\mbox{Re}\ \Big<\left[\partial^i_vD({\bf v},{\bf v'})\right]\partial^j_{v'}D({\bf v'},{\bf v}) \Big>_{x_2}
\right.\\\left.
+\Big<D({\bf v},{\bf v'})D({\bf v'},{\bf v})\partial^i_v\partial^j_{v'}  Q({\bf x},{\bf y},{\bf v'},{\bf v})\Big>_{x_2}\Big|_{\substack{{\bf x}={\bf v}\\{\bf y}={\bf v'}}}
-\frac{1}{N^2_c}\partial^i_v\partial^j_{v'}
O({\bf x},{\bf v},{\bf v'},{\bf x},{\bf y},{\bf v'},{\bf v},{\bf y})\Big>_{x_2}\Big|_{\substack{{\bf x}={\bf v}\\{\bf y}={\bf v'}}}
\right\}\ .
\eea

When all the pieces are put together, we do recover the small-$x$ limit of the TMD formula \eqref{eq:tmd-gg}. In particular, the combination of gluon TMDs which is sub-leading in $N_c$, ${\cal F}_{gg}^{(4)}+{\cal F}_{gg}^{(5)}-2{\cal F}_{gg}^{(3)}$, does come out of the octupole term.

\section{Lattice simulation of the JIMWLK equation}
\label{sec:lattice}

\subsection{Algorithmic implementation}

The JIMWLK renormalization group equation has been solved on a lattice
in transverse space with a Langevin process in the space of Wilson
lines by Rummukainen and Weigert \cite{Rummukainen:2003ns}. We use here the
efficient left-right factorization algorithm introduced in \cite{Lappi:2012vw}:
\begin{align}
  \label{ALGO}
  U_{\b{x}}(s+\delta s) &= \exp\lp{-\sqrt{\delta s}\sum_{\b{y}} U_{\b{y}}(s) 
    \lp{\v{K}(\b{x}-\b{y})\cdot\v{\xi}(\b{y})}U_{\b{y}}^{\dagger}(s)} U_{\b{x}}(s)
  \exp\lp{\sqrt{\delta s}\sum_{\b{y}}\v{K}(\b{x}-\b{y})\cdot\v{\xi}(\b{y})}
\end{align}
where
\be
s=\frac{\alpha_s}{\pi^2}\ y\ ,\quad\mbox{with}\quad y=\ln\left(\frac{x_0}{x_2}\right)\ .
\ee
This expresses the evolution in $x_2$, or rapidity $y$, of each Wilson line $U_{\b{x}}(s)$, at fixed coupling, in terms only of the
convolution of the BFKL kernel $\v{K}(\b{x}-\b{y})$ with a white noise $\v{\xi}(\b{x})$ in the Lie algebra. Such an algorithm brings about a reduction factor $4/(N_c^2+3)$ in computing time for the gauge group $SU(N_c)$. Since these update algorithms ignore terms of order 4 and
higher in the Langevin step $\sqrt{\delta s}$, it is sufficient to expand the exponentials up to order 3 and project the expansions onto
the group.

$x_0$ denotes the starting point of the evolution, at $y=0$. For the sake of comparison we choose to construct the initial
configurations on the lattice exactly as described in detail by Lappi
in Ref.~\cite{Lappi:2007ku}. We generate directly in Fourier space of a
periodic square lattice of size $L$, a translation-invariant color
field distribution in the $SU(3)$ Lie algebra with the variance of the
discretized McLerran-Venugopalan (MV) model,
\begin{align}
  \label{MV} 
  \vev{\h{A}^a_{\b{k}}\h{A}^{b\star}_{\b{k}'}} =
  g^4\mu^2\delta^{ab}\f{\delta_{\b{k},\b{k'}}}
  {\lp{4\sum_i \sin^2\f{\omega k_i}{2}}^2}\,,
  \qquad\omega=\f{2\pi}{L}\,, \qquad 0\leq k_1,\,k_2 < L\,.
\end{align}
(We denote integer lattice momentum components by $k_i$ and momentum
components in the Brillouin zone by $p_i=\omega k_i$). The zero mode
is of course removed. Then the Wilson lines are approximated by a
product of $N_y=100$ infinitesimal adjoint fields with variance
$\propto g^4\mu^2/N_y$.

By construction, in the absence of an infrared regulator $m$ in
\eqref{MV}, only the dimensionless parameter $g^2\mu$ (in lattice
spacing units $a=1$) appears in the numerical calculation of initial
configurations. Therefore lines of constant ''physics'' of the MV
model are obtained by letting $L\r\infty$ and $g^2\mu\r 0$ so that
$g^2\mu L$ remains constant. Indeed the color fields of the MV model
are classical fields and all correlators of Wilson lines fall on
universal curves at fixed $g^2\mu L$. Physical lengths are defined by
fixed values along these universal curves. In practice it is most
convenient to define correlation lengths from the values of the
dimensionless dipole correlator of two Wilson lines in the fundamental
representation, normalized to unity at $\b{x}=\b{y}$,
\begin{align}
  \label{Dipole} 
  C(|\b{x}-\b{y}|) = \f{1}{N_c}\vev{\tr U_{\b{x}}^{\dagger}U_{\b{y}}} \,.
\end{align}
We shall follow the standard convention of defining a
Gaussian-like correlation length $R_s$ by \D{C(R_s)=e^{-1/2}}.

From a numerical standpoint the optimal value of $R_s$ in lattice
spacing units should minimize both discretization errors and
finite-size effects.  Discretization errors are best exhibited by the
lattice derivative $D_i$ which is defined as a central derivative with
discretization errors of order $a^2$ in terms of the forward and
backward derivatives,
\begin{align}
  \label{Di}
D_i U_{\b{x}}= \frac12 \left(U_{\b{x}+\h{\imath}}-U_{\b{x}-\h{\imath}} \right)\, .
\end{align}
With this definition, the continuum momentum $p_i$ is replaced by the
well-nown sine function on the lattice,
\begin{align}
  \label{Pi}
  p_i\quad\longrightarrow\quad\h{p}_i = \sin p_i\,,
\end{align}
which shows that it is unreliable to study lattice momenta beyond
$\f{\pi}{4}$ in the Brillouin zone. The same bound $L/8$ should be
appplied in coordinate space to the tails of Wilson line
correlators. Hence a safe upper bound to the initial correlation
length $R_s$ is given by
\begin{align}
  \label{Bound}
  2R_s \lesssim \f{L}{8}\,.
\end{align}
On the other hand the initial correlation length should be as large as
possible since the renormalization group evolution in rapidity drives
correlation lengths to zero.

The gluon distributions listed in Eqs.~\eqref{eq:fqg1}, \eqref{eq:fgg3} and \eqref{eq:fqg2}-\eqref{eq:fgg6} are two-point functions defined
as products of various traces which must be evaluated component-wise in
order to express them as scalar convolution products which can be
calculated using a discrete fast Fourier transform algorithm. For
instance the distribution $\FC_{gg}^{(3)}(x_2,p_\perp)$, which is the Weizs\"acker-Williams gluon distribution already studied
numerically in \cite{Dumitru:2015gaa}, can be evaluated very efficiently as
\begin{align}
  \label{WW}
  \FC_{gg}^{(3)}(x_2,p_\perp) = \f{8\pi}{g^2}
  \sum_{i=1}^2\sum_{a,b=1}^{N_c}
  \vev{\left|\int\f{d^2x}{(2\pi)^2}e^{-i\b{p}_\perp\cdot\b{x}}
      \lp{U_{\b{x}}^{\dagger}\p_i U_{\b{x}}}_{ab} \right|^2}_{x_2}\ .
\end{align}
The distribution $\FC_{gg}^{(4)}$ can be written similarly as a sum of
$2N_c^2$ convolution products, the distributions $\FC_{qg}^{(2)}$,
$\FC_{gg}^{(1)}$ and $\FC_{gg}^{(2)}$ as a sum of $2N_c^4$ convolution
products, whereas the distributions $\FC_{gg}^{(5)}$ and
$\FC_{gg}^{(6)}$, more costly, require $2N_c^6$ convolution
products. Most convolution products necessitate, as \eqref{WW}, a
single Fourier transform, except for the distributions
$\FC_{gg}^{(2)}$ and $\FC_{gg}^{(5)}$, where two Fourier transforms
are needed because of the reshuffling of indices between the two
factors.

The operator $U^{\dagger}\p_\mu U$ which appears in Eq.~\eqref{WW}
shows up within the expression of most gluon distributions listed in \eqref{eq:fgg3}-\eqref{eq:fgg6},
except $\FC_{gg}^{(2)}$, $\FC_{gg}^{(4)}$ and the usual
dipole operator $\FC_{qg}^{(1)}$. Care must be exercised in the
lattice discretization of this operator to keep its anti-hermiticity
property of the continuum, namely
\begin{align}
  \label{AH}
  \lp{U^{\dagger}\p_\mu U}^{\dagger} = -U^{\dagger}\p_\mu U\ .
\end{align}
We have already described the lattice discretization \eqref{Di} of the
derivative operator. The anti-hermiticity property \eqref{AH} is
enforced on the lattice by the substitution
\begin{align}
  \label{AHL}
  U_{\b{x}}^{\dagger}\p^\mu U_{\b{x}}\quad\longrightarrow\quad
  A^{\mu}_{\b{x}} = \f{1}{2}\lp{U^{\dagger}_{\b{x}}D^\mu U_{\b{x}} -
    \lp{U^{\dagger}_{\b{x}}D^\mu U_{\b{x}}}^{\dagger}}
\end{align}
We have a similar discretization for the operator $\lp{\p_\mu U}U^{\dagger}$
which enters $\FC_{gg}^{(4)}$.

\subsection{Data Analysis}

All numerical measurements of the gluon distributions reported in this
work have been performed on the lattice size $L=1024$. A statistical
sample of 64 independent initial configurations has been generated
with the smallest rounded parameter value, $g^2\mu=0.03$, compatible
with the upper bound for the correlation length:
\begin{align}
  \label{Rs}
  \vev{R_s} = 65.8\pm 0.3\,.
\end{align}
All averages or data points displayed in this work have error bars
which have been determined from the JIMWLK evolution of this random
sample with a Langevin step $\delta s = 0.0001$, which is still
adequate for our lattice size.

Before turning to the presentation of results, one must tackle a
problem which becomes manifest when statistical errors are smaller
than systematic errors. The phenomenon is illustrated by
Fig.~\ref{fig:QG1}.
\begin{figure}
  \begin{center}
    \includegraphics[width=0.9\textwidth]{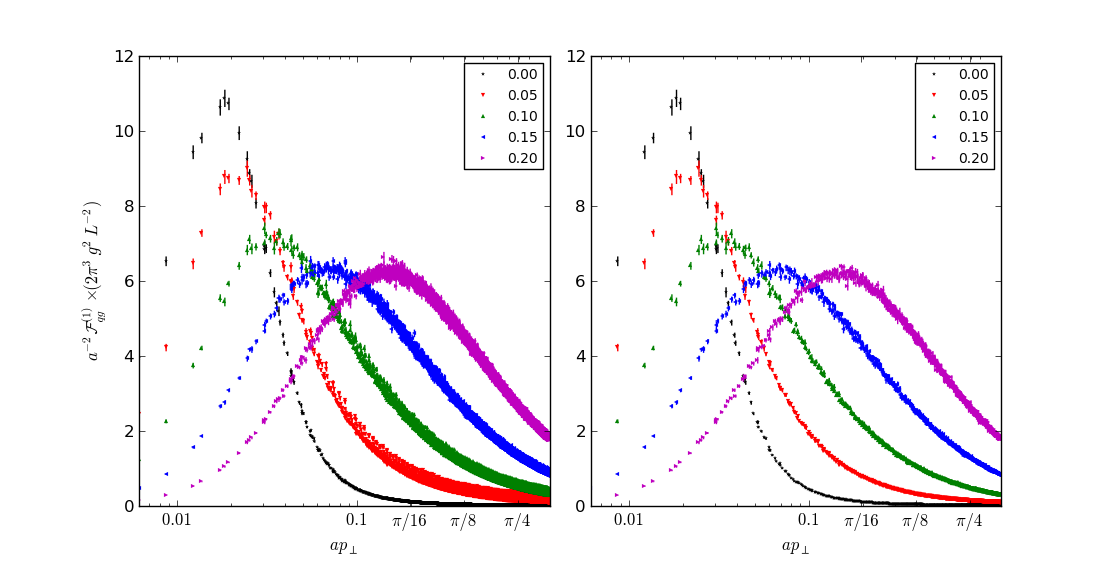}
  \end{center}
  \caption{JIMWLK evolution of the gluon distribution $\FC^{(1)}_{qg}$
    in momentum space at selected values of $\alpha_s y/\pi^2$, without any cut (left) and with the
    near-$O(2)$-invariant selection \eqref{eq:cut} (right). The linear
    vertical scale is rescaled by the factor $2\pi^3g^2L^{-2}$. Only raw
    lattice data points with error bars are displayed; there is no
    interpolating line.}
  \label{fig:QG1}
\end{figure}
Whereas the JIMWLK evolution of the dipole correlator at fixed
coupling is very smooth in coordinate space, in qualitative agreement
with the results of Refs.~\cite{Rummukainen:2003ns,Lappi:2007ku}, the evolution in Fourier
space is afflicted at small rapidity by huge discretization errors
which are negligible in the initial configurations at $y=0$. Indeed
different lattice points with the same $k^2$ need not have the same
correlator value. This peculiarity produces a characteristic spread in
raw lattice data. The scatter of data points in the initial MV
configurations is reduced because of the simple sine-squared momentum
dependence \eqref{MV} of gluon propagators. A probable explanation
of the increase of the spread by the JIMWLK evolution is the scale
invariance of the BFKL kernel in the continuum,
\begin{align}
  \label{eq:kernel}
  \v{K}(\v{r}) = \f{\v{r}}{r^2}\,,
\end{align}
which is certainly badly broken by a naive discretization on a
periodic lattice. We have taken the periodicity into account by
modifying the BFKL kernel as in \cite{Rummukainen:2003ns}. The statistical noise
increases with the rapidity evolution and the orbit structure becomes
less visible at high rapidity.

The issue of lattice artifacts in momentum space is well-known in
numerical studies of the gluon propagator in lattice QCD. There, it is
due to the breaking of rotational invariance by the square
lattice. The lattice correlators depend not only upon the O(2)
invariant $k^2=k_1^2+k_2^2$, but also on the other independent
invariant $k^4\equiv k_1^4+k_2^4$ of the symmetry group $D_4$ of the
square. The action of the dihedral group $D_4$ on lattice points
generates orbits which are characterized by both invariants and
partition the lattice. Since the invariant $k^4$ is not present in
the continuum limit, in principle one should extrapolate the data to $k^4\r 0$.
There are sophisticated techniques to perform such an extrapolation
in four dimensions \cite{deSoto:2007ht}. In two dimensions there are not enough
orbits to apply the full machinery.

But there is a standard recipe which is particularly effective in two
dimensions. It consists of performing a cut to the data which removes
momenta with the highest $k^4$ at fixed $k^2$. Since $k^4$ is
minimized at fixed $k^2$ when $k_1=k_2$, it is sufficient to remove
momenta when $|k_1-k_2|$ exceeds a certain threshold. In all the data
analyses below, we choose to keep the momenta $k=(k_1,k_2)$
with
\begin{align}
  \label{eq:cut}
  |k_1-k_2| \leq 5\,.
\end{align}
The cut is chosen as an empirical compromise between the requirements
of smoothness of data points at high $k^2$ and their paucity at low
$k^2$. The near-$O(2)$ invariant subset for the gluon distribution
$\FC^{(1)}_{qg}$ is exhibited in Figure~\ref{fig:QG1}~(right).

\section{Results for the gluon TMDs}
\label{sec:results}

\subsection{Geometric Scaling}

\begin{figure}[t]
  \begin{center}
    \includegraphics[width=\textwidth]{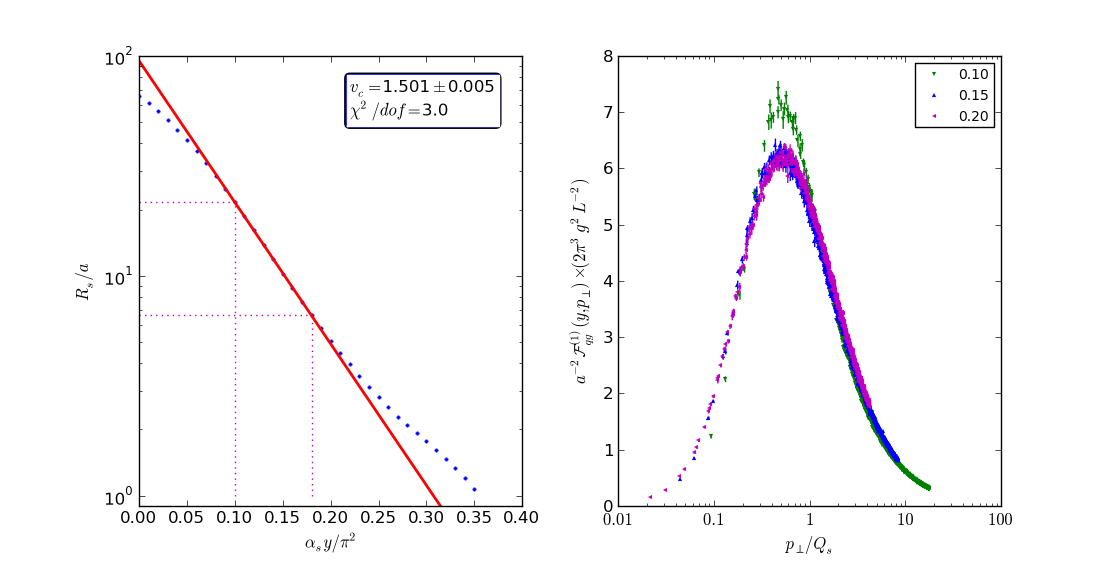}
  \end{center}
  \caption{(left) Rapidity evolution of the correlation length $R_s$;
    dotted lines mark the boundaries of the linear regression fit
    displayed as a solid line. (right) Gluon distribution
    $\FC^{(1)}_{qg}$ in momentum space as a function of the scaling
    variable $p_{\perp}/Q_s(y)$ inside the geometric scaling
    window. The linear vertical scale is rescaled by the factor
    $2\pi^3g^2L^{-2}$.}
    \label{fig:Rs_scaling}
\end{figure}

From general considerations \cite{Munier:2003vc}, one expects that various gluon TMDs ${\cal F}(y,p_\perp)$, evolved in rapidity with the JIMWLK equation, should reach an asymptotic regime called ``geometric scaling'', which is independent of the lattice size or initial correlation length. This means that the TMDs are functions of $p_\perp R_s(y)$ only, as opposed to $p_\perp$ and $y$ separately. Following \cite{Rummukainen:2003ns}, we display in
Fig.~\ref{fig:Rs_scaling} the logarithmic variation of $R_s(y)$ with rapidity and Table~\ref{tab:Rs} lists selected values.

Usually geometric scaling is characterized by looking for a plateau in
the instantaneous evolution speed
\begin{align}
  \label{eq:speed}
  v_c = -\f{\p\ln R_s(y)}{\p(\alpha_s y)}\,.
\end{align}
But numerical derivatives are rather noisy and we prefer to perform a
linear regression on $\ln R_s(y)$ which gives a better statistical
estimation. We can identify clearly in
Fig.~\ref{fig:Rs_scaling}\,(left) a window of pretty good geometric
scaling in the interval $0.10\lesssim \alpha_s y/\pi^2\lesssim 0.20$
around the inflexion point. A presence of pre-asymptotic terms and
residual systematic errors much larger than the statistical errors can
be inferred from the sensitivity of the $\chi^2/dof$ to the fitting
window. In particular it can be seen that the scaling regime is
sensitive to lattice spacing effects as soon as $R_s\lesssim
4a$. Therefore one should avoid matching curves at smaller $R_s$. The
lattice artifacts entail that the widths $\Delta_sy$ of geometric
scaling windows depend logarithmically on the lattice size,
\begin{align}
  \alpha_s\Delta_sy \approx \f{1}{v_c}\ln\lp{\f{R_s(y_s)}{4}}
  \lesssim \f{1}{v_c}\ln\lp{\f{L}{64}} \,,
\end{align}
where $y_s$ is the lower bound of a scaling window (and depends, like
$v_c$, on the precise definition of $C(R_s)$).

\begin{table}[t]
\centering
  \begin{tabular}{|c|c|c|}
    \hline $\alpha_s y/\pi^2$ & $R_s(y)$ \\
    \hline
    0. & 65.79(28) \\
    0.05 & 41.31(17) \\
    0.10 & 21.58(6) \\
    0.15 & 10.23(3) \\
    0.20 & 5.07(1) \\
    \hline
  \end{tabular}
  \caption{Rapidity evolution of the correlation length $R_s$.}
  \label{tab:Rs}
\end{table}

The approximate scaling is confirmed in
Fig.~\ref{fig:Rs_scaling}\,(right) which displays the gluon
distribution $\FC^{(1)}_{qg}$ at different rapidities inside the
geometric scaling window as a function of the scaling variable
$p_{\perp}/Q_s(y)$, where the saturation scale $Q_s$ is related to $R_s$
according to the convention
\begin{align}
  \label{eq:Qs}
  Q_s(y) = \f{\sqrt{2}}{R_s(y)}\,.
\end{align}
Figure~\ref{fig:TMDs_scaling} shows that, for all other gluon TMDs we
have measured, effective geometric scaling does hold pretty well around
the saturation scale over a window in momentum space which spans roughly
one order of magnitude. We have checked that the distributions
$\FC^{(3)}_{gg}$ and $\FC^{(4)}_{gg}$ are identical within statistical
errors and we display only one of the two. In principle, the agreement
could still be slightly improved since the saturation scale can be
adjusted for every distribution as it may differ from \eqref{eq:Qs} by an
overall factor in the geometric scaling window.

Since the saturation scale $Q_s(y)$ increases exponentially with the
rapidity, there are two distinct regimes of scaling violations in
momentum space. Near the upper bound of the geometric scaling window
in rapidity space, for $\alpha_s y/\pi^2\simeq 0.2$, $Q_s(y)$ is large and the violations of scaling at
large $p_\perp$ occur at lattice momenta above a few units of the saturation scale, where lattice spacing effects become sizable. By
the same token finite-size effects are small in this regime and geometric scaling seems to hold down to rather low $p_{\perp}\sim
0.1\,Q_s(y)$.

On the other hand, near the lower bound of the geometric scaling
window, for $\alpha_s y/\pi^2\simeq 0.1$, the saturation scale $Q_s(y)$ becomes small and
scaling violations show up at momenta below the saturation scale,
where finite-size effects are important. For the same reasons lattice
spacing effects are small in this regime and geometric scaling may
hold up to $p_{\perp} \sim 20\,Q_s(y)$.

In \cite{Dominguez:2011gc}, it was conjectured that the color quadrupole, and subsequently
the Weizs\"acker-Williams gluon distribution, should obey the same geometrical
behavior as the dipole gluon distribution. We have shown, for the first time, that
this conjecture holds not only for the Weizs\"acker-Williams gluon distribution but also for all the other
gluon TMDs found in forward di-jet production.

Geometric scaling has already been investigated qualitatively from the
solution of the JIMWLK equation with a running coupling for the color
dipole \cite{Lappi:2011ju}, in coordinate and momentum spaces, as well as for
the color quadrupole \cite{Dumitru:2011vk} in coordinate space. Smoothness of our
near-$O(2)$-invariant dataset, at fixed coupling, ensures that we
should be able to extract anomalous dimensions quantitatively (a task
which is routinely performed in lattice calculations of the gluon
propagator on much smaller lattices) and test theoretical models.

\begin{figure}
  \begin{center}
    \includegraphics[width=\textwidth]{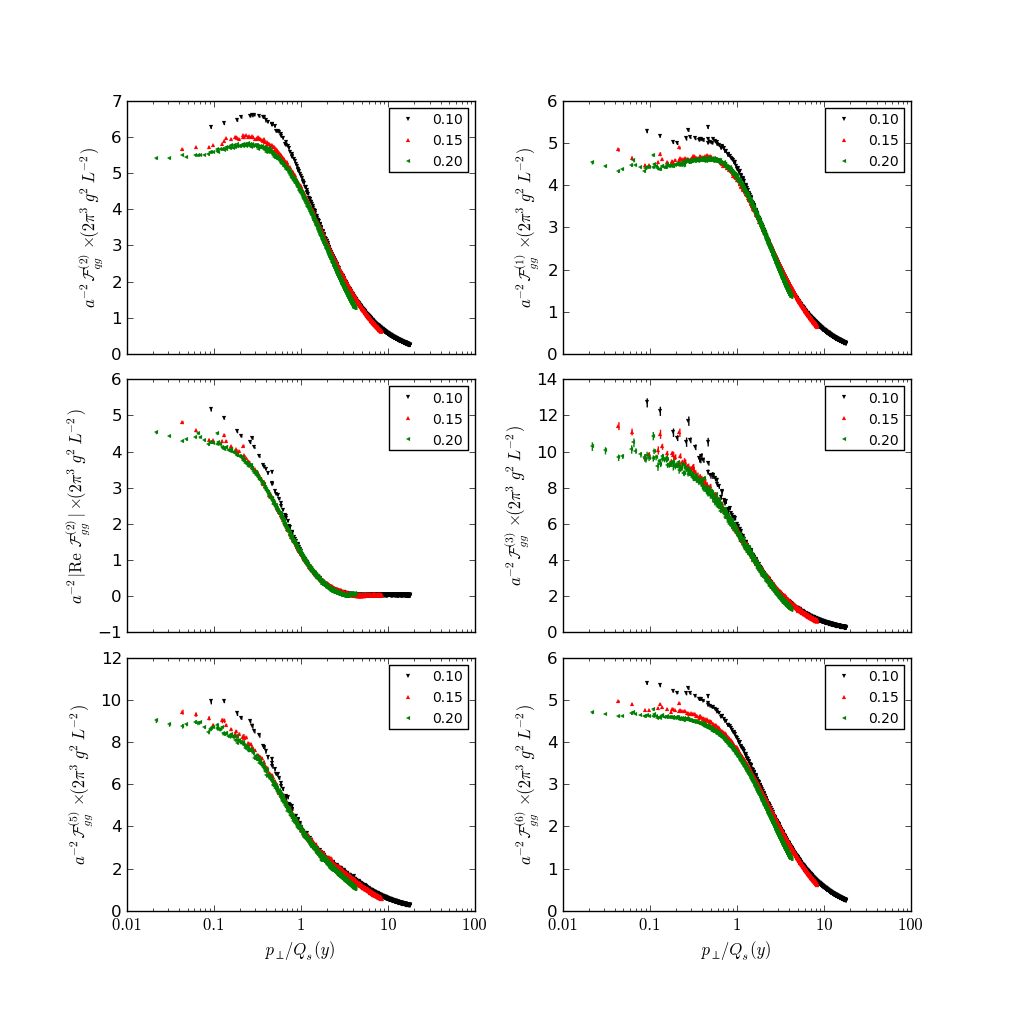}
  \end{center}
  \caption{Transverse momentum dependence of gluon TMDs at selected values of
    $\alpha_sy/\pi^2$ within the geometric scaling window. All linear
    vertical scales are rescaled by the factor $2\pi^3g^2L^{-2}$. The
    saturation scale $Q_s(y)$, extracted from the fundamental dipole amplitude \eqref{Dipole}, is the same in all subplots.}
  \label{fig:TMDs_scaling}
\end{figure}

\subsection{High-$k_t$ behavior}

\begin{figure}
  \begin{center}
    \includegraphics[width=0.8\textwidth]{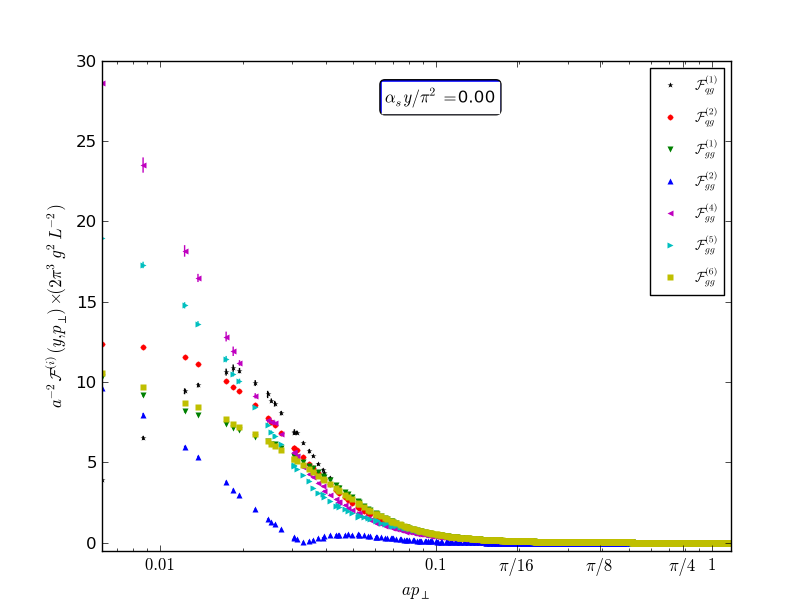}
  \end{center}
  \caption{Transverse momentum dependence of gluon TMDs in the initial MV
    configurations. The linear vertical scale is rescaled by the
    factor $2\pi^3g^2L^{-2}$. The logarithmic momentum scale is in inverse
    lattice spacing units. The label $\mathcal{F}^{(2)}_{gg}$ is a shorthand for
    $\vert\rm{Re}\,\mathcal{F}^{(2)}_{gg}\vert$, and $\FC^{(3)}_{gg}$, not shown, is identical to $\FC^{(4)}_{gg}$.}
  \label{fig:TMDs_0}
\end{figure}

Recently, it was shown that in the initial MV configurations at $y=0$, all the gluon TMDs must have a universal
$1/p^2_\perp$ behavior at large $p_\perp$ \cite{vanHameren:2016ftb,Petreska:2015rbk}, expect for ${\cal F}_{gg}^{(2)}$ which should vanish faster.
Fig.~\ref{fig:TMDs_0} displays the gluon TMDs in the MV model calculated on the lattice, and Table~\ref{tab:Qs} lists the best
parameters of a power-law behavior of each gluon distribution separately within the asymptotic window in momentum space,
\begin{align}
  \label{eq:universality}
  \lp{2\pi^3g^2}(La)^{-2}\FC(p_\perp) = 2\pi N_c
  \lp{\f{\o{Q}^2_s}{p^2_\perp}}^{\gamma}\,,\qquad
  \f{\pi}{16}\leq ap_\perp \leq \f{\pi}{4}\,.
\end{align}
The values for the anomalous dimension are pretty close to the
theoretical value and vindicate our claim that our analysis is
not only qualitative but also quantitative. The asymptotic window
looks small on a logarithmic scale but each fit has $\gtrsim$ 300
independent degrees of freedom for a lattice size $L=1024$, after
the cut \eqref{eq:cut}. As a matter of fact, statistical errors
in Table~\ref{tab:Qs} are much smaller than the $\sim 3\%$ systematic
errors in the power law. The residual systematic errors are partly
due the approximations in the lattice discretization \eqref{MV} of
the MV model and partly to the discretization of gluon TMDs.
The parameter $N_y=100$ could be fine-tuned to minimize some of
these discretization errors.

A byproduct of this analysis is that it is possible to define a
natural saturation scale $\o{Q}_s$ in the MV model:
\begin{align}
  \label{eq:saturation}
  a\o{Q}_s = 0.015(1)\,
\end{align}
which is to be compared to the Gaussian-like definition from the fundamental dipole amplitude \eqref{Dipole}, i.e. Eq.~\eqref{eq:Qs} at $y=0$, whose value is $aQ_s \approx 0.0215$.

\begin{table}[b]
  \centering
  \begin{tabular}{|c|c|c|c|}
    \hline
    TMD & $a\o{Q}_s$ & $\gamma$ & $\chi^2/dof$ \\
    \hline
    $\FC^{(1)}_{qg}$ & 0.014(1) & 0.99(1) & 1.3 \\
    $\FC^{(2)}_{qg}$ & 0.015(1) & 1.04(1) & 1.1 \\
    $\FC^{(1)}_{gg}$ & 0.015(1) & 1.05(1) & 0.9 \\
    $\FC^{(3)}_{gg}$ & 0.015(1) & 1.03(1) & 1.1 \\
    $\FC^{(5)}_{gg}$ & 0.015(1) & 1.05(1) & 1.1 \\
    $\FC^{(6)}_{gg}$ & 0.015(1) & 1.05(1) & 1.1 \\
    \hline
  \end{tabular}
  \caption{Asymptotic saturation scale $a\o{Q}_s$ in inverse lattice
    spacing units and power law $\gamma$ in the MV model.}
  \label{tab:Qs}
\end{table}

After some evolution, the high-p$_{\perp}$ behavior is best elucidated by looking at the top plot of 
Fig.~\ref{fig:TMDs_evolved}, which shows the gluon TMDs for $\alpha_s y/\pi^2=0.1$, after enough evolution to have reached
the scaling regime, but not too much so that the high-p$_{\perp}$ tails of the gluon distributions stay
within the accessible momentum range on the lattice. We observe that the initial properties survive,
meaning that that all gluon distributions fall onto a universal curve at high-p$_{\perp}$, except $\rm{Re}\,\FC^{(2)}_{gg}$ which vanishes ($\FC^{(3)}_{gg}$ is not displayed in the figure but is identical to $\FC^{(4)}_{gg}$). What changes after some evolution, as expected, is the power-law behavior of that universal tail, which becomes less steep than $1/p_\perp^2$. A detailed study of that anomalous dimension in the geometric scaling window is left for future work.

\begin{figure}
  \begin{center}
    \includegraphics[width=0.8\textwidth]{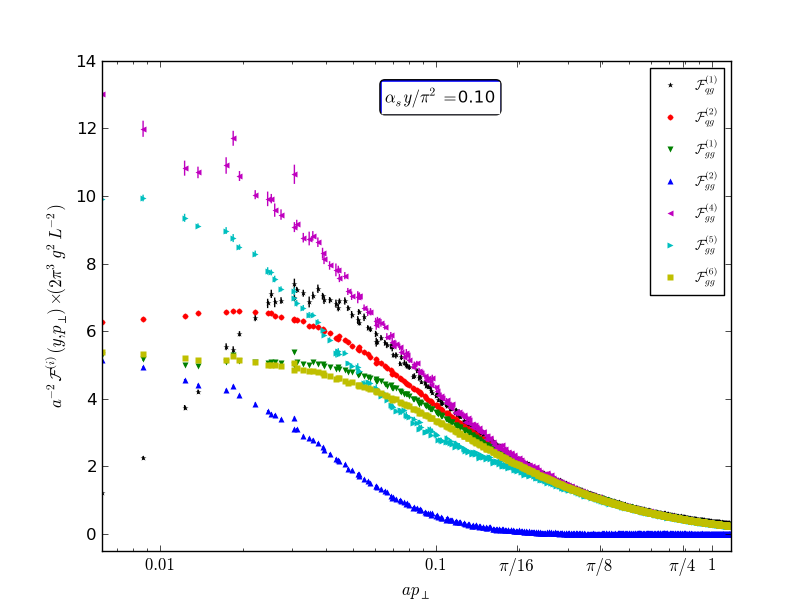}
    \includegraphics[width=0.8\textwidth]{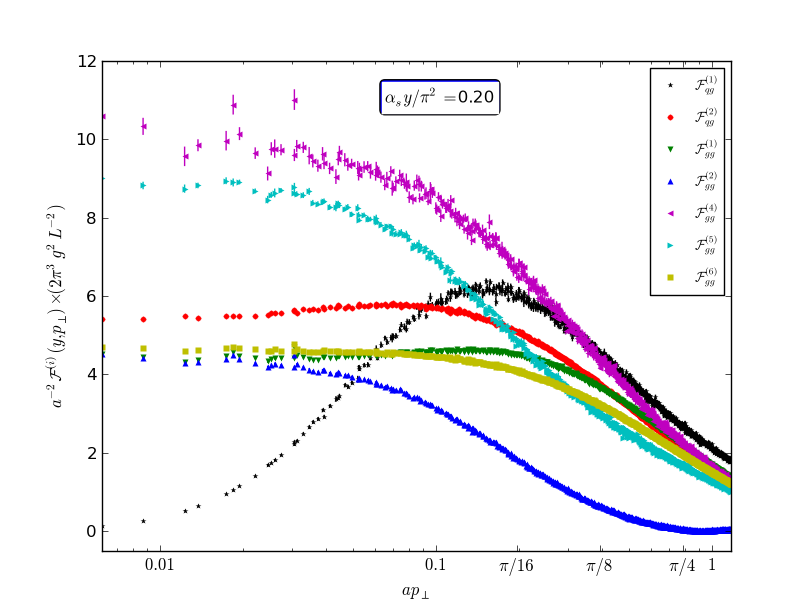}
  \end{center}
  \caption{Momentum dependence of gluon TMDs near the lower bound (top) and upper bound (bottom) of
    the geometric scaling window. The logarithmic momentum scale is in inverse lattice spacing units. The label
    $\mathcal{F}^{(2)}_{gg}$ is a shorthand for $\vert\rm{Re}\,\mathcal{F}^{(2)}_{gg}\vert$.}
  \label{fig:TMDs_evolved}
\end{figure}

For completeness, we also show in the bottom of Fig.~\ref{fig:TMDs_evolved}, the gluon TMDs after further evolution at $\alpha_sy/\pi^2=0.2$, where the high-$p_\perp$ tail has disappeared from the accessible momentum range of our analysis. However, what is interesting is that conversely, this gives us better access into the saturation regime at low $p_\perp$, where the various gluon TMDs are very different from each other, and where the process dependence of TMDs is the most relevant. Indeed, while at high-p$_{\perp}$ the process dependence of gluon TMDs may be safely ignored (expect for peculiar operator definitions like that of
${\cal F}^{(2)}_{gg}$), it certainly cannot be inside the saturation regime.

Note finally that combining the two plots of Fig.~\ref{fig:TMDs_evolved}, we have enough information at small and large transverse momentum to provide parameterizations, for all those unpolarized gluon TMDs, which could be used in phenomenological studies. For instance, in a future work we plan to extend the work of \cite{vanHameren:2016ftb}, in which the large-$N_c$ limit was assumed, and the Balitsky-Kovchegov equation \cite{Balitsky:1995ub,Kovchegov:1999yj} used to evolve the gluon TMDs.

\section{Conclusions}

We have studied the process dependence of unpolarized gluon TMDs at small-$x$, in the CGC framework. As a testing ground, we considered
forward di-jet production in p+p or p+A collisions, a process in which kinematics impose a dilute-dense asymmetry, $x_1\sim 1$ and $x_2\ll 1$, that in turn
allows to employ TMD distributions for the small-$x_2$ target only. We investigated what happens when the large gluon density of the target reaches the
saturation regime, and how the various process-dependent gluon TMDs are affected by non-linear effects when the transverse momentum becomes of the
order of the saturation scale $Q_s(x_2)$, or below.

For nearly back-to-back jets, when the total transverse momentum of the jet pair $k_t$ is much smaller than the individual jet transverse momenta $\sim P_t$,
the process can be described in the TMD factorization approach, and the cross section is given by formulae \eqref{eq:tmd-qg}-\eqref{eq:tmd-gg}, which involve
eight distinct gluon TMDs, functions of $x_2$ and $k_t$, with different operator definitions: \eqref{eq:tmdqg1}-\eqref{eq:tmdgg6}. We showed that in the
small-$x$ limit, these can be simplified and written in terms of CGC correlators of Wilson lines: \eqref{eq:fqg1} for the "dipole" gluon distribution, \eqref{eq:fgg3}
for the Weizs\"acker-Williams distribution, and \eqref{eq:fqg2}-\eqref{eq:fgg6} for the six others.

In the small-$x$ limit, forward di-jet production can be described in the CGC framework as well, without imposing any particular
ordering of the di-jet transverse momentum scales. The cross-section is then given by formulae \eqref{eq:cgc-qg}, \eqref{eq:cgc-qqbar}
and \eqref{eq:cgc-gg}, which involve correlators of up to eight Wilson lines, all of which sit at different transverse positions. We showed
that in the $k_t\ll P_t$ limit, these CGC expressions simplify to coincide with formulae \eqref{eq:tmd-qg}-\eqref{eq:tmd-gg}, giving complete
agreement between the CGC and TMD frameworks, in their overlapping domain of validity, without resorting to the large-$N_c$ limit.
In the CGC the various process-dependent gluon TMDs emerge as different Wilson line correlators: \eqref{eq:fqg1}, \eqref{eq:fgg3}
and \eqref{eq:fqg2}-\eqref{eq:fgg6}.

This allows their evaluation from the fixed-coupling JIMWLK equation, including the full small-$x$ QCD evolution with non-linear corrections. We use the standard two-dimensional lattice formulation in terms of a Langevin process in the space of Wilson lines. We obtain that at large transverse momentum, $k_t\gg Q_s(x_2)$, the process dependence essentially disappears, in the sense that all the gluon TMDs fall onto a universal curve, except for $\rm{Re}\,\FC^{(2)}_{gg}$ which vanishes faster, as shown in Fig.~\ref{fig:TMDs_evolved} (top plot). This feature of the MV model which we used as an initial condition is preserved by the evolution, with the difference that the power-law fall-off of the gluon TMDs with $k_t$ evolves from $1/k_t^2$ (a theoretical result which we checked here numerically) to a less steep power: $1/k_t^{2\gamma}$ with $1/2<\gamma<1$.

By contrast, the process dependence of the gluon TMDs is most relevant at small transverse momentum $k_t\leq Q_s(x_2)$: as shown in Fig.~\ref{fig:TMDs_evolved} (bottom plot), the various distributions are very different from each other. However, it is important to point out that at small-x, in the CGC framework, these differences are fully under control, hereby restoring universality: potential information extracted from a particular process, for one gluon TMD, can be consistently fed into the others. Indeed, for phenomenology it will be important for instance to take into account running-coupling corrections, and the parameters related to its unknown behavior in the non-perturbative region will need to be adjusted to fit data. Some of the gluon TMDs we have considered here also enter in the formulation of other processes such as di-jet production in deep-inelastic scattering or photon-jet production in hadronic collisions, and the results we have obtained within this work are also applicable in those cases.

Most importantly, we have observed that all the TMD gluon distributions reach a geometric scaling regime after some evolution, i.e. they become functions of 
$k_t/{Q}_s(y)$. A precise determination of the properties of this regime, such as the $y$ dependence of the saturation scale and of the anomalous dimension
$\gamma$, whose existence is manifest from Fig.~\ref{fig:TMDs_evolved}, is outside the scope of the present study.
But such a determination is quite possible, especially if one considers that reproducing the numerical data generated in this analysis requires a computing time four orders of magnitude less than typical ab-initio lattice QCD calculations. Moreover, numerical analyses should be as precise for JIMWLK evolutions supplemented with running-coupling corrections \cite{Lappi:2012vw} or collinear resummations \cite{Hatta:2016ujq}. For genuine higher-order corrections \cite{Balitsky:2013fea,Kovner:2013ona}, that will depend on whether or not the Langevin description still holds.

Finally, it would be interesting to extend our analysis to the case of polarized gluon TMDs, as very little is known on what happens to their process dependence at small $x$. First works considering polarized hadrons or nuclei have appeared recently \cite{Kovchegov:2015zha,Kovchegov:2015pbl,Boer:2016xqr}, for the "dipolar" gauge link structure of ${\cal F}^{(1)}_{qg}$, but to our knowledge none of the other structures have been looked at in this context. Note that linearly-polarized gluons in unpolarized hadrons are also of interest, and in our framework at small-x, it is straightforward to obtain all the corresponding TMDs. This is done by projecting the transverse Lorentz indices of the field correlators onto a different structure than $\delta_{ij}$.

\section*{Acknowledgments}

We are grateful to Tuomas Lappi and Heribert Weigert for discussions and clarifications on their JIMWLK lattice implementations. CM would also like to thank Jian Zhou for motivating him to study the CGC/TMD equivalence beyond the large-$N_c$ limit. EP acknowledges support from: European Research Council grant HotLHC ERC-2011-StG-279579; Ministerio de Ciencia e Innovacion of Spain under project FPA2014-58293-C2-1-P; Xunta de Galicia (Conselleria de Educacion) within the Strategic Unit AGRUP2015/11.

\end{document}